\documentclass[11pt]{article}
\usepackage{fullpage}

\usepackage{etoolbox}

\makeatletter
\patchcmd{\@makecaption}
  {\parbox}
  {\advance\@tempdima-\fontdimen2} 
  {}{}
\makeatother

\usepackage{latexsym}
\usepackage[pdftex]{graphicx}
\usepackage[pdftex]{color}
\usepackage[normalem]{ulem} 
\RequirePackage[OT1]{fontenc}
\RequirePackage{amsmath,amssymb}
\RequirePackage{natbib}
\RequirePackage[colorlinks,citecolor=blue,urlcolor=blue]{hyperref}
\usepackage{epsfig,xcolor}
\usepackage{comment}

\usepackage{calc}
\usepackage{fancyhdr}
\usepackage{ifthen}
\usepackage{tabularx}

\def\spacingset#1{\renewcommand{\baselinestretch}%
{#1}\small\normalsize} \spacingset{1}

\usepackage{xr}
\makeatletter
\newcommand*{\addFileDependency}[1]{
  \typeout{(#1)}
  \@addtofilelist{#1}
  \IfFileExists{#1}{}{\typeout{No file #1.}}
}
\makeatother

\newcommand{\mX}{\ensuremath{\mathcal X}}
\newcommand{\mL}{\ensuremath{\mathcal L}}
\newcommand{\mE}{\ensuremath{\mathbb E}}
\newcommand{\mP}{\ensuremath{\mathbb P}}
\newcommand{\mR}{\ensuremath{\mathbb R}}
\newcommand{\vmu}{\ensuremath{{\vec \mu}}}

\newcommand{\vD}{\ensuremath{\vec D}}
\newcommand{\II}{\ensuremath{\mathbb I}}
\newcommand{\vu}{\ensuremath{\vec u}}
\newcommand{\vv}{\ensuremath{\vec v}}
\newcommand{\vLambda}{\ensuremath{\vec \Lambda }}
\newcommand{\vh}{\ensuremath{\vec h }}
\newcommand{\vx}{\ensuremath{\vec X}}
\newcommand{\vE}{\vec E}

\newcommand{\rhadd}[1]{\textcolor{red}{#1}}

\newtheorem{definition}{Definition}[section]

\newtheorem{proposition}{Proposition}[section]
\numberwithin{equation}{section}
\newtheorem{prop}{Proposition}[section]
\newtheorem{thm}{Theorem}[section]
\newtheorem{lem}{Lemma}[section]
\newtheorem{coro}{Corollary}[section]
\newtheorem{ass}{Assumption}[section]

\begin{document}
\title{ Optimal and Maximin Procedures for Multiple Testing Problems }

\author{Saharon Rosset  and Ruth Heller and Amichai Painsky and Ehud Aharoni}


\maketitle

\begin{abstract}
Multiple testing problems (MTPs) are a staple of modern statistical analysis. The fundamental objective of MTPs is to reject as many false null hypotheses as possible (that is, maximize some notion of power), subject to controlling an overall measure of false discovery, like family-wise error rate (FWER) or false discovery rate (FDR). In this paper we 
provide generalizations to MTPs of the optimal Neyman-Pearson test for a single hypothesis. We show that for simple hypotheses, for both FWER and FDR and relevant notions of power, finding the optimal multiple testing procedure can be formulated as infinite dimensional binary programs and can in principle be solved for any number of hypotheses. We also characterize maximin rules for complex alternatives, and demonstrate that such rules can be found in practice, leading to improved practical procedures compared to existing alternatives that guarantee strong error control on the entire parameter space. We demonstrate the usefulness of these novel rules for identifying which studies contain signal in numerical experiments as well as in  application to clinical trials with multiple studies. In various settings, the increase in power from using optimal and maximin procedures can range from 15\% to more than 100\% .    
\end{abstract}


\section{Introduction}\label{sec-Introduction}

Hypothesis testing is a fundamental component of scientific research. Having methods for assuring validity of results and controlling false discoveries is critical, to avoid a situation where ``most published research is false'' as has often been claimed in recent years \cite{Ioannidis05, Simmons11}. However, it is arguably just as critical that these methods not be over-conservative, since this limits the ability of scientists to make true discoveries. Methods that control false discovery while being as liberal as possible yield higher statistical power, and facilitate more discoveries, thus are highly desirable for scientific use.

Major advances have been made since the late 90's of methods for control of false discovery rate (FDR) in lieu of more traditional and conservative measures like familywise error rate
(FWER). The Benjamini-Hochberg (BH) suite of FDR control methods gained tremendous influence due to their ability to increase power and make more discoveries while provably controlling a relevant
measure of false discovery.  However, such methods are not appropriate  when false discoveries cannot be tolerated, as, e.g., is the case in clinical trials with multiple studies/endpoints/subgroups. Hence development of new and powerful MTPs for controlling either FDR or FWER remains of prime importance.

 The problem of developing ``most powerful'' MTPs for these problems have thus far been addressed in the literature to a limited extent and mostly under major simplifying assumptions  \citep{spjotvoll72,Westfall98, Storey07, Sun07,Lehmann05,  Dobriban15}.
For example, although we know that BH guarantees FDR control, it is not the ``most powerful'' method for controlling it.

In a classic hypothesis testing problem, we are given null and alternative hypotheses, and we wish to find good statistical tests for this problem. A good test is expected to be valid and have the desired probability of rejection under the null model, while being powerful and having a high probability of rejection under the alternative.
When the hypotheses are both simple and fully specify the distribution of the data, the Neyman-Pearson (NP) Lemma characterizes the most powerful test at every given level, as rejecting for high values of the likelihood ratio.

This ``most powerful test'' problem can be viewed as an optimization problem, where every point in sample space has to be assigned to {\em reject} or {\em non-reject} regions, in a manner that maximizes the expected rejection under the alternative distribution, subject to a constraint on its expectation under the null. When the sample space is infinite (such as a Euclidean space), this is an infinite dimensional binary optimization problem, whose optimal solution happens to have the simple structure characterized by NP.

When moving from testing a single hypothesis to multiple testing scenarios, several complications are added. First, there is no longer a single universally accepted definition of false discovery. Given a rejection policy, denote the (random) number of rejected hypotheses by $R$, and the number of falsely rejected hypotheses (true nulls) by $V$.  Two commonly used  measures of false discovery, which we denote generically by $Err$, are:
\begin{eqnarray*}
\mbox{FWER: } \mP(V>0); \quad
\mbox{FDR: }  \mE\left( \frac{V}{R} ; R>0\right).
\end{eqnarray*}
The problem is especially challenging since we require {\em strong} $Err$ control, i.e., $Err\leq \alpha$ for all possible configurations of null and non-null hypotheses and parameter values. This is the commonly used error control requirement for MTPs, and it expresses the fact that the true configuration is unknown.

Second, there is no longer a single notion of power. For example, we may seek a test which maximizes the expected number of rejections if all nulls are false, or one which maximizes our chance of correctly rejecting a single false null, or we may want to maximize the expected number of true rejections under some (prior, estimated or known) distribution on the percentage of false nulls, as in the Bayesian approach to FDR \citep{Genovese02,Efron10}. The chosen definition should capture the true ``scientific'' goal of the testing procedure and the type of discoveries we wish to make.

However, once we choose a false discovery criterion  and a power criterion,  we can write the problem of finding the optimal test as an optimization problem.  


 Our first challenge in this paper is to develop theory and algorithms for such {\em Optimal Multiple Testing} (OMT) procedures for simple hypotheses, and investigate the implications that our results have on design of practical multiple testing procedures, and we address this in \S~\ref{Sec:main-res}. Our main result   is that the goal of finding OMT procedures is attainable in principle for any exchangeable MTP. The key observation towards this result is that like in single hypothesis testing, the key test statistic is the likelihood ratio of each test, and the optimal policy is a step-down procedure on this statistic. We prove optimality and derive the resulting algorithm using Lagrange theory.  We apply our algorithms to find OMT procedures for testing three  normal means with strong FWER or FDR control. The resulting OMT procedures are much more powerful than relevant alternatives. 
 
 However, in realistic testing problems the assumption of simple hypotheses is usually not realistic. In particular, accounting for complexity of alternative hypotheses (potential discoveries) is critical, in assuring that strong control is guaranteed for any combination of relevant alternatives.  In the classic results, the NP theory is expanded to notions of optimality for complex or one-sided alternatives, like uniformly most powerful (UMP) tests or {\em maximin} tests maximizing the minimal power among the relevant alternatives \citep{LehmannRomano05}.
 
 To address the challenge of optimally dealing with complex alternatives in MTPs, we define in \S~\ref{Sec:maximin} a maximin formulation of multiple testing with complex alternatives.  For this setting we derive sufficient conditions for existence of maximin solutions, and formulate the resulting algorithm. We demonstrate that these sufficient conditions hold in interesting examples, such as controlling FWER or FDR for testing normal means, allowing us to find maximin procedures for complex alternatives.  In \S~\ref{subsec-cochrane} we demonstrate the utility of maximin OMT procedures with FWER control in subgroup analysis of randomized clinical trials. 
 
 It is important to note that the maximin formulation relaxes the requirement of exchangeability --- the actual distribution can be different for each of the testing problems where the null does not hold. When the sufficient conditions hold, the maximin solution is valid across all possible combinations of such alternatives, i.e., it provides strong control of the error rate. Moreover, the maximin solutions are fundamentally different than existing alternatives. For example, all procedures for FWER control are closed testing procedures \citep{Goeman20}, but the vast majority of existing  methods can be expressed as applying the closure principle \citep{Marcus76} to global tests, while our approach cannot, with the desirable result of significant power increase while still guaranteeing strong FWER control.

Table \ref{tab-taxonomy} presents the taxonomy of optimality definitions and our original solutions in the context of MTPs. 

 \begin{table}
\caption{\label{tab-taxonomy} Optimality definitions in single testing, parallels for MTPs, and our contributions. }
\begin{tabular}{|c|c|c|}
  \hline
 & Simple alternative & Complex alternative \\
 \hline 
 Single & Most powerful (NP) & UMP / maximin \\
 hypothesis& & \citep{LehmannRomano05} \\
 \hline 
 Multiple & {\bf OMT}  & {\bf maximin}  \\
 testing & (\S~\ref{Sec:main-res}) & (\S~\ref{Sec:maximin}) \\
 \hline
\end{tabular}
\end{table}

The optimal rejection regions presented in \S~\ref{Sec:main-res} and \S~\ref{Sec:maximin} can have counter-intuitive shapes, in particular display {\em non-monotone} behaviors, where smaller p-values can lead to fewer rejections. \rhadd{In \S~\ref{sec:weakmon} } we discuss and interpret these interesting behaviors noting that they are typical for low-power settings, but tend to vanish as power increases. We also offer in \S~\ref{sec:weakmon} an extension which enforces properly defined monotonicity and demonstrate that we can still solve the maximin problem formulation with the added constraints, and that the loss of power from enforcing it is typically small. 


Optimality results for MTPs available in the literature typically start from seeking the optimal procedure for a  restricted class of decision rules (e.g., single step procedures), and within this restricted class they provide procedures with optimal properties, see \cite{spjotvoll72, Westfall98, Lehmann05, Dobriban15} for results for FWER control and  \cite{Genovese02, Storey07, Sun07, Castro17, Finner09} for results for  FDR control. For the specific setting covered by the ``two-group model'', where  the test statistics are generated from a mixture distribution of null and non-null densities, \cite{Sun07, Xie11} provide optimal procedures for marginal FDR control, and we provide the optimal procedure for FDR control in \cite{Heller19}. A notable exception which considers all possible decision rules, like our approach, is \cite{Rosenblum14}, which 
used a discrete optimization approximation (as compared to our infinite optimization approach) to estimate the OMT procedure for two independent hypotheses with normal distributions. In \S~\ref{Sec:related work} we compare and contrast our optimality results with the above previous work.

Our key contributions, OMT solutions for exchangeable MTPs and maximin solutions for non-exchangeable MTPs (with or without enforcing monotonicity of the rejection region), lead to novel MTPs with non-negligible power improvements over existing alternatives. The new formulations are conceptually important, since they are the first (as far as we know) that do not restrict the class of decision rules in which to search for the optimal solution. However,   
the computational complexity of the resulting algorithms is high, and therefore the dimension (number of hypotheses $K$) of problems that can be solved is limited with current algorithms. We demonstrate here solutions for up to $K=3$ hypotheses. We show  that this is already useful for important applications like those in  \S~\ref{subsec-cochrane}. In \S~\ref{Sec:discuss} we present some directions for increasing the dimension where solutions can be found.

\section{OMT procedures for exchangeable simple hypotheses}
\label{Sec:main-res}

\subsection{Problem formulation and notation} 
We start by assuming that we have $K>1$ identical hypotheses testing problems, and that we have a  sample for testing each of the K null hypotheses. 
For the $i$th hypothesis testing problem, using data vector $\vx_i$, let $\mL_0(\vx_i)$ and $\mL_1(\vx_i)$ be the null and non-null likelihood, and $\Lambda(\vx_i)= \frac{\mL_1(\vx_i)}{\mL_0(\vx_i)}$ the likelihood ratio (LR).

We denote the true states of all $K$ tests   by the fixed (yet unknown) vector $\vec h\in\{0,1\}^K$, where, for the $k$th hypothesis testing problem, $h_k=1$ if  the alternative hypothesis is true and $h_k=0$ if the null hypothesis is true.   
Let $\mL_{\vh} (\vx)$ be the likelihood of $\vx = (\vx_1,\ldots, \vx_K)$ when the true configuration is $\vec h$, and $\vLambda(\vx) = (\Lambda(\vx_1), \ldots, \Lambda(\vx_K))$ be the vector of likelihood ratios. 
We denote by $\vec h_L,\; 0\leq L\leq K$ the special configuration with the first $L$ nulls being false, and the rest true:
$$
\vec h_L = (\underbrace{1,1,...,1}_L,\underbrace{0,....,0}_{K-L})^t.
$$

Let $S_K$ be the permutation group for $K$ elements. For $\sigma \in S_K$, 
$\sigma(\vx_1,\ldots, \vx_K) $ and $\sigma(\vh)$ are the corresponding permutation of the data vector $\vx$ and hypothesis state vector $\vh$. We denote by $\sigma_{ij}\in S_K$ the permutation that interchanges entries $i$ and $j$ only. For example, $\sigma_{12}(\vx) = (\vx_2, \vx_1,\vx_3,\ldots, \vx_K)$, $\sigma_{12}(\vh) = (h_2, h_1,h_3,\ldots, h_K)$, and $\vLambda(\sigma_{12}(\vx)) = (\Lambda(\vx_2), \Lambda(\vx_1),\Lambda(\vx_3), \ldots, \Lambda(\vx_K))$. 

If $\vx_1,\ldots,\vx_K$ are mutually independent,  the likelihood of $\vx = (\vx_1,\ldots, \vx_K)$ is $\mL_{\vh}(\vx) = \prod_{k=1}^K \mL_{h_k}(\vx_k)$. Consider two coordinates $i$ and $j$ such that  $\Lambda(\vx_i)>\Lambda(\vx_j)$ and $h_i\geq h_j$. Then $  \mL_{\vh}(\vx) \geq \mL_{\sigma_{ij}(\vh)}(\vx)$   with strict inequality only if $h_i=1$ and $h_j=0$, since then $$ \frac{ \mL_{\vh}(\vx)}{ \mL_{\sigma_{ij}(\vh)}(\vx)} = \frac{\mL_1(\vx_i)\mL_0(\vx_j)}{\mL_0(\vx_i)\mL_1(\vx_j}= \frac{\Lambda(\vx_i)}{\Lambda(\vx_j)}>1.$$ It turns out that this order relation between the LRs and the joint likelihood  is enough in order to prove that the OMT procedure will reject first the largest LR statistics. 

We shall prove that the OMT procedure rejects first the largest LR statistics in a slightly more general setting, where $\vx_1,\ldots,\vx_K$ need not be mutually independent but they need to satisfy the following two conditions (which are clearly satisfied if $\vx_1,\ldots,\vx_K$ are mutually independent). 

\begin{ass} \label{ass:exch}
{\bf $\vec h$ - Exchangeability:} The $K$ tests are $\vec h$-exchangeable if
$$
\mL_{\vec h}(\vx) = \mL_{\sigma(\vec h)} \left[\sigma(\vx)\right]\;,\; \forall \vx,\; \sigma \in S_K.
$$
As a consequence $\vx_i$ and $\vx_j$ necessarily have the same distribution if $h_i=h_j$. 
\end{ass}

\begin{ass} \label{ass:monoLR}
{\bf Arrangement increasing:}   For any data $\vx$, state vector $\vh,$ and permutation $\sigma$ that satisfies  $\Lambda(\vx_k) \geq \Lambda(\sigma(\vx)_k), \forall k$  with $h_k=1$, we have:
$$
\mL_{\vec h}(\vx) \geq \mL_{\vec h}(\sigma(\vx)).
$$
In words, if the LRs of the non-null hypotheses in $\vLambda(\vx)$ are larger than those in $\vLambda(\sigma(\vx))$
 (and consequently, the opposite holds for the LRs of null hypotheses), then the joint density at $\vx$ is higher that at $\sigma(\vx)$.
\end{ass}

We call Assumption \ref{ass:monoLR} {\em Arrangement Increasing} because if $\left(\Lambda(\vx_i)-\Lambda(\vx_j)\right)(h_i-h_j)\leq 0$, then by interchanging entries $i$ and $j$ in the vector $\vx$ to form the vector $\sigma_{ij}(\vx)$, we have the relation $\mL_{\vec h}(\vx)\leq \mL_{\vec h}(\sigma_{ij}(\vx))$.
In \S~\ref{supp-DistArrInc}
we provide examples for which $\vx_1,\ldots,\vx_K$ are not mutually independent but Assumption \ref{ass:monoLR} is satisfied.

A testing policy has to make a decision which hypotheses are rejected at each possible data vector with the binary decision  function $\vec D:\mX \rightarrow \{0,1\}^K$, where $\mX$ is the range of $\vx$. Given the $\vec h$-exchangeability assumption, we limit our interest to functions that are symmetric, i.e.,  whose decision does not depend on the order of the hypotheses\footnote{Given the definitions that follow, we can in fact prove that optimal procedures are indeed symmetric without assuming this form, if we allow randomized policies. For simplicity we choose to state this as a requirement here.}:
\begin{definition} \label{def:sym}
A decision function $\vec D: \mX \rightarrow \{0,1\}^K$ is {\em symmetric} if
$$
\sigma\left(\vec D(\vx)\right) =  \vec D\left(\sigma(\vx)\right) \;,\; \forall \vx\in \mX,\; \sigma \in S_K.
$$
\end{definition}

To formulate the OMT problem as an optimization problem we need to select the false discovery criterion we wish to control, and the power function we wish to optimize.  For power many notions can be considered, as discussed in \S~\ref{sec-Introduction}. In this paper we limit our consideration to the following options that serve well our purpose of demonstrating the utility of our approach:
\begin{eqnarray}
&& \Pi_{any}(\vec D)= \mP_{\vec h_K} (R(\vec D)(\vx) > 0) \label{pow1}\\
&& \Pi_L(\vec D) = \mE_{\vec h_L} [D_1(\vx)+...+D_L(\vx)]/L\;,\; 1\leq L \leq K. \label{pow2}
\end{eqnarray}
In words, $\Pi_{any}$ is the probability of making any discoveries if all alternatives are true, and it was discussed for example in \cite{Lehmann05, Bittman09}. $\Pi_L$ is the average power (also known as total power, \citealp{Westfall01}), and it seeks to maximize the expected number of true rejections given that $L$ nulls are false.
Note that although calculated assuming the density is $f_{\vec h_L}$, due to the $\vec h$-exchangeability assumption and symmetry requirement on $\vec D$, the value of $\Pi_L$ would be the same if the expectation is calculated relative to any other configuration of $L$ false nulls.

Given a selected power measure $\Pi$ and false discovery measure to control $Err$, we can write the OMT problem of finding the optimal test subject to strong control as an infinite dimensional binary program, where the optimization is over the value of the function $\vec D$ at every point in the cube:
\begin{eqnarray}
\label{opt} \max_{\vec D: \mX \rightarrow \{0,1\}^K,\mbox{symmetric}} && \Pi(\vec D)\\
\mbox{s.t. } &&  Err_{\vec h_L} (\vec D) \leq \alpha\;,\; 0\leq L < K. \nonumber
\end{eqnarray}
We denote the optimal solution to this problem (assuming it exists) by $\vec D^*$. We have only $K$ constraints and not $2^K-1$ due to $\vec h$-exchangeability and symmetry.

Several aspects of this optimization problem appear to make it exceedingly difficult to solve:
\begin{enumerate}
\item[1.] The optimization is over an infinite number of variables
\item[2.] This is an integer (binary) optimization problem, which can be impractical to solve even in finite dimensional cases.
\end{enumerate}

\subsection{The ordering of OMT procedures by their likelihood ratio statistics}

Our first key theoretical result is that an optimal solution to problem  (\ref{opt}) necessarily rejects first the largest likelihood ratio statistics.
\begin{thm}\label{lemma-weakmonotone}
Under Assumptions \ref{ass:exch},\ref{ass:monoLR}, for any of our considered power and level criteria, the optimal symmetric solution $\vec D^*$ is a function of the LR statistics, and it satisfies the {\em LR-ordering property}, i.e., it rejects the largest LRs at every realization $\vx$:
$$ \Lambda(\vx_i) \geq \Lambda(\vx_j) \Rightarrow   D^*_i(\vx) \geq D^*_j(\vx).$$
\end{thm}

Proofs are  supplied in \S~\ref{supp:proofs} in the supplementary material. Note that the LR-ordering property holds more generally, specifically also for $K$ simple non-exchangeable hypotheses, if $\vx_1,\ldots,\vx_K$ are mutually independent, see \S~\ref{supp-LRorderinggeneral} for details. 

Since Theorem \ref{lemma-weakmonotone} states that rejections are based on the LR statistics, we can in fact limit the definition of $\vec D$ to consider only the ordered LR statistics. The  $p$-value of a LR statistic is necessarily a non-increasing function of the LR, so without loss of generality we move to consider the LR $p$-values: for realized LR $\Lambda_i$ for the $i$ hypothesis testing problem,  $u_i = \int I\{\vx_i: \frac{\mL_1(\vx_i)}{\mL_0(\vx_i}\geq \Lambda_i  \}\mL_0(\vx_i)d\vx_i$. The null density of $u_i$ is uniform on (0,1), and we denote its alternative density  by $g(\cdot)$. 

Our MTP thus formulated is to test the following family of $K$ hypotheses: 
 \begin{eqnarray*}
 H_{0k} \ : \ U_k \sim U(0,1); \quad
 H_{Ak} \ : \ U_k \sim g.
 \end{eqnarray*}

 Given the symmetry of $\vD$ and the fact that $u_1,\ldots,u_K$ are sufficient statistics for the MTP, we  limit the definition of $\vec D$ to consider only the ``lower corner'' set $Q = \{u: 0\leq u_1 \leq u_2 \leq\ldots\ \leq u_K\leq 1\}$,
 and extend it to $[0,1]^K$ through the symmetry. Throughout our discussion we limit our attention to functions $\vec D$ that are Lebesgue measurable on $Q$  or $[0,1]^K$ (note that $\vec D$ is also bounded by definition and so integrable).

The LR-ordering  result in Theorem \ref{lemma-weakmonotone} implies that  on the ``lower corner'' set  $Q$, $\vec D^*$ can be characterized via
 $
 k^*(\vec u) = \mbox{max} \{k\leq K : D^*_k(\vec u) =\ 1\},
 $
 the ``last and largest'' $p$-value rejected by $\vec D^*$ at $\vec u\in Q$.
 Given this solution on $Q$ we can extend it to $[0,1]^K$ using the symmetry property:
 $\vec D^*(\vec u) = \sigma_{\vec u}^{-1}\left(\vec D^*(\sigma_{\vec u}(\vec u))\right)$, where $\sigma_{\vec u}$ is the sorting permutation for $\vec u$, so that $\sigma_{\vec u}(\vec u) \in Q $ is the order statistic of $\vec u$.

\subsection{The Linear representation of the objectives and constraints}

Once we limit our discussion to functions $\vec D$ that have this structure, we can simplify the mathematical description of the objective and constraints of our optimization problem. 
For this purpose, we add the following notation. Let $f_{\vh}$ denote the joint density of the $K$ $p$-values when the true configuration is $\vh$, e.g., $f_{\vh_L}$ is the density for the configuration with the first $L$ nulls being false. 

Taking into account $\vec h$-exchangeability and symmetry, the L-expected power can be written as a linear functional of $\vec D$ on $Q$:
    \begin{eqnarray}
   \Pi_L(\vec D) &=& \mE\left(\frac{\sum_{k=1}^K D_k(\vec u)h_k}{L} \right)= \int_{[0,1]^K}  f_{\vec h_L}(\vec u) \sum_{k=1}^L D_k(\vec u) d\vec u/L = \nonumber \\ &=&  L!(K-L)!\int_{Q}  \sum_{i \in \binom{K}{L}} f_i(\vec u) \sum_{k \in i} D_k(\vec u)  d\vec u/L, \label{obj}
    \end{eqnarray}
 where $i$ indexes the set of all subsets of size $L$; the notation $k\in i$ is shorthand that the $k$th null is set to false by the $i$th configuration; $f_i(\cdot)$ is the density under the configuration of $L$ false nulls indexed by $i$. The second equality follows since without loss of generality we assume the first $L$ hypotheses are the the non-null hypotheses. The third equality follows by expressing the integral on the  ``lower corner'' set $Q$.
 
    Second, the LR-ordering property  of the optimal solution (Theorem~\ref{lemma-weakmonotone})  is sufficient to simplify $\Pi_{any}$ to a linear functional too:
    \begin{equation}
    \label{obj-any} \Pi_{any}(\vec D) = K! \int_{Q}  D_1(\vec u) f_{\vec h_K}(\vec u) d\vec u.
     \end{equation}

    Moving to the constraints, symmetry and $\vec h$-exchangeability allow us to write the  constraints of Problem (\ref{opt}) in the form :
    $$
    FWER_{\vec h_L}(\vec D) = \int_{\vec u \in [0,1]^K} \mathbb I \left\{V(\vec D(\vec u)) > 0 \right\}  f_{\vec h_L}(\vec u) d\vec u \leq \alpha,\; 0 \leq L < K .
    $$
    A similar expression can be written for FDR.
By Theorem~\ref{lemma-weakmonotone}, we can rewrite these $K$ constraints as linear functionals of the decision function $\vec D$ on $Q$ for both FWER and FDR:
    \begin{eqnarray}
     \label{FWER-const} &&
          FWER_L(\vec D) = L!(K-L)!\int_Q \sum_k D_k(\vec u)
    \sum_{ \substack{i \in \binom{K}{L}\\ \bar{i}_{min} = k} } f_i(\vec u) d\vec u , \\
\label{FDR-const}     && FDR_L(\vec D) = L!(K-L)! \int_Q \sum_k D_k(\vec u)
    \sum_{i \in \binom{K}{L}} f_i(\vec u) r_{ki}  d\vec u.
    \end{eqnarray}
    where $\bar{i}_{min}$ is the minimal element not in the $i$'th configuration of false nulls (that is, the true null hypothesis with the smallest index), and $r_{ki}$ is the difference in false discovery proportion (FDP) if we reject the $k$ versus $k-1$ smallest $p$-values, i.e.,
 
    $$ r_{ki} = \frac{\left|\{1,\ldots\,k\} \cap i^c\right|}{k} - \frac{\left|\{1,\ldots\,k-1\} \cap i^c\right|}{k-1}, $$
    where $i^c$ denotes the actual set of true nulls in the configuration indexed by $i$ (and we also assume $0/0 = 0$). See  
    \S~\ref{supp-FDRexpression}  for  the derivation of $FDR_L(\vec D).$

Taking all of these together we conclude that for any combination of objective of the form $\Pi_{any},\;\Pi_L$ and  strong control of FWER or FDR, we can rewrite Problem (\ref{opt}) as an infinite linear binary program on the set $Q$, with a linear objective and $K$ linear constraints.

\subsection{The OMT procedures}

We aim to solve the following optimization problem: 

\begin{eqnarray}
 &\max_{\vec D: Q \rightarrow \{0,1\}^K}&  \int_{Q}  \left(\sum_{i=1}^K a_i(\vec u) D_i(\vec u)\right) d\vec u   \label{eq-primal}\\
&\mbox{s.t. }&   \int_{Q} \left(\sum_{i=1}^K b_{L,i}(\vec u)D_i(\vec u)\right) d\vec u \leq \alpha\;,\; 0\leq L < K. \nonumber,
\end{eqnarray}
where $a_i,\; b_{L,i},\;i=1,\ldots,K,\;L=0,\ldots,K-1$ are fixed non-negative real functions over $Q$, and are linear combinations of the density functions $\left\{f_{\vec h}:\vec h\in\{0,1\}^K\right\}$, with non-negative coefficients that depend on the specific choice of $\Pi, Err$, as shown in Eqs. (\ref{obj})-(\ref{FDR-const}).

From the LR-ordering property (Theorem \ref{lemma-weakmonotone}) it follows that   for $\vec u \in Q$ we need to consider only policies  in the range   $\mathcal D= \lbrace (0,\ldots, 0), (1,0,\ldots,0), \ldots, (1,\ldots,1,0), (1,\ldots, 1)\rbrace,$ instead of $\{0,1\}^K.$

The Lagrangian is 
\begin{eqnarray}
L(\vD,  \mu)& = &  \int_{Q}  \left(\sum_{i=1}^K a_i(\vec u) D_i(\vec u)\right) d\vec u+\sum_{L=0}^{K-1}\mu_L\times \left \lbrace \alpha -  \int_{Q} \left(\sum_{i=1}^K b_{L,i}(\vec u)D_i(\vec u)\right) d\vec u \right \rbrace \nonumber \\
& = & \sum_{L=0}^{K-1}\mu_L\times\alpha + \int_{Q} \left \lbrace\sum_{i=1}^KD_i(\vec u)\left( a_i(\vec u)- \sum_{L=0}^{K-1}\mu_L \times b_{L,i}(\vec u)\right)\right \rbrace d\vec u
 \label{eq:Lagrangian} 
\end{eqnarray}
for $ \mu = (\mu_0, \ldots, \mu_{K-1})$  with nonnegative entries (i.e., $\mu_L\geq 0, L = 0, \ldots, K-1$).

Let $f^*$ denote the optimal (maximum) value of problem \eqref{eq-primal}. Clearly, 
$$f^*\leq \max_{\vD : Q\rightarrow \mathcal D}L(\vD,  \mu).$$

Let
$ R^{\mu}_k(\vu) = a_k(\vu) - \sum_{L=0}^{K-1}\mu_L \times b_{L,k}(\vu) $. 

It is easy to see that  for every $\vu$, $$\max_{\vD(\vu)\in \mathcal D}\sum_{k=1}^K D_k(\vu)R^{\mu}_k(\vu)=\max\left \lbrace \max_{l=1,\ldots,K} \sum_{k=1}^l R^{\mu}_l(\vu) ,0 \right \rbrace$$
achieved for the following policy: 
\begin{eqnarray}
D^\mu_1(\vu) &=&  \II\left\{\cup_{l=1}^K \left(\sum_{k=1}^l R^{\mu}_k(\vu) > 0\right)\right\} \label{Dmu1}\\
D^\mu_i(\vu) &=&  \II\left\{\left(\vD^\mu_{i-1}(\vu)=1\right) \cap \cup_{l=i}^K \left(\sum_{k=i}^l R^{\mu}_k(\vu) > 0\right)\right\},\; i=2,\ldots,K. \label{Dmu2}
\end{eqnarray}
 
 It thus follows that $$\vD^{\mu}(\vu) = \arg \max_{\vD : Q \rightarrow \mathcal D}L(\vD, \mu). $$

\begin{proposition}\label{prop-zeroduality}
If $\exists \ \mu^* \geq \vec 0 $ such that $\vD^{\mu^*}:\mR^K\rightarrow  \mathcal D$ satisfies for $L=0,\ldots,K-1$:
\begin{eqnarray}
&&\left(\mu^*_L \geq 0 \cap \int_{Q} \left(\sum_{i=1}^K b_{L,i}(\vec u)D^{\vmu^*}_i(\vec u)\right) d\vec u=\alpha \right) \cup \label{eq:mucond}\\ &&\;\;\;\;\;\;\;\;\;\left(\mu^*_L = 0 \cap \int_{Q} \left( \sum_{i=1}^K b_{L,i}(\vec u)D^{\vmu^*}_i(\vec u)\right) d\vec u\leq \alpha \right), \nonumber
\end{eqnarray}
 then $\vD^{\mu^*}$ is an optimal policy for Problem \eqref{eq-primal}.
\end{proposition} 
{\bf Proof:}
$L(\vD^{\mu^*}, \mu^*)= \int_{Q}  \left(\sum_{i=1}^K a_i(\vec u) D^{\mu^*}_i(\vec u)\right) d\vec u$, so $L(\vD^{\mu^*}, \mu^*)\leq f^*$ since $\vD^{\mu^*}$ is feasible (for $L\in \{0,\ldots,K-1\}$ and $\mu_L^*\geq 0$ it satisfies $\int_{Q} \left(\sum_{i=1}^K b_{L,i}(\vec u)D^{\mu^*}_i(\vec u)\right) d\vec u \leq \alpha$). 
On the other hand, $f^*\leq L(\vD^{\mu}, \mu) \ \forall \ \mu\geq \vec 0$. Therefore, $f^* = L(\vD^{\mu^*}, \mu^*)$ and the optimal solution is achieved at $\vD^{\mu^*}$.

Thus, a complete algorithm for solving our OMT problems involves:
\begin{enumerate}
\item An approach for searching the space $(\mathbb R^+ \cup \{0\})^K$ of possible $\vmu$ vectors for a solution $\vmu^*$.
\item An approach for (exact or numerical) integration , to calculate
$$\int_{Q} \left(\sum_{i=1}^K b_{L,i}(\vec u)D^\vmu_i(\vec u)\right) d\vec u$$ for any $\vmu$ vector and asses the error relative to the conditions on $\vmu^*$.
\end{enumerate}



In \S~\ref{supp:alg} we demonstrate a detailed derivation of the formulas and resulting algorithm for a specific instance of the general problem:  Maximizing $\Pi_3$ for $K=3$ independent tests under FDR control.

A solution exists  under mild ``non-redundancy" conditions. We make the following assumption: 
\begin{ass} \label{ass:nonred}
The set of density functions $\left\{f_{\vec h}: \vec h\in\{0,1\}^K \right\}$ is {\em non-redundant}, i.e., any non-trivial linear combination of the $2^K$ density functions is non-zero almost everywhere on $[0,1]^K$:
$$\sum_{\vec h} \gamma_{\vec h} f_{\vec h}(\vec u)\neq 0   \textrm{ almost everywhere for any fixed vector } \gamma_{\vec h}\in\mR, \ \sum_{\vec h} |\gamma_{\vec h}|>0.$$
\end{ass}
This assumption is mild given the highly non-linear nature of the functions $f_{\vec h}$ in typical applications (as in our examples below).

\begin{proposition}\label{prop-existence}
Under Assumption \ref{ass:nonred}, the optimal solution $\vmu^*$ exists.  
\end{proposition}
 

\subsection{Example applications: $K=3$ independent normal means} 

We now utilize our results to illustrate the potential power gain from using OMT procedures as well as the potential insight  gained from examining the optimal solutions.

We consider tests of the form $H_{0k}: X \sim N(0,1)$ vs $H_{Ak}: X \sim N(\theta,1)$ for $k=1,2,3$ with $\theta<0$, where all test statistics are independent. The power functions we consider are $\Pi_{\theta, 3}(\vec D)$, the {\em average power} when all three nulls are false, and $\Pi_{\theta, any}(\vec D)$, the probability of making at least one rejection when all three nulls are false, which we term {\em minimal power}.


For strong FWER control, we demonstrate the potential power gain over the popular  sequentially rejective procedure of Holm \citep{Holm79}, henceforth Bonferroni-Holm.
In \cite{Lehmann05} the step-down Sidak procedure (which performs at each step Sidak's test \citealp{Sidak67}) was shown to maximize certain aspects of power\footnote{The specific aspect of power considered in \cite{Lehmann05} was the following: the minimal probability of at least one rejection, among all configurations with $K$ exchangeable non-null hypotheses with signal $\theta\geq \epsilon$.}, among all monotone rejection policies (defined in \S~\ref{Sec:related work}). Since for $\alpha=0.05$,  the difference of this policy  from Bonferroni-Holm is negligible (less than $10^{-5}$ in average or minimal power for our experiments below),  we only compare henceforth with the widely used  Bonferroni-Holm.

%

Table \ref{tab-power-FWER} shows the  power comparison for various values of $\theta$. The power of Bonferroni-Holm is  smaller than the power of the OMT policy for $\Pi_{\theta,3}$  by more than 20\%. However, the average power of the OMT policy for $\Pi_{\theta,any}$ can be lower than that of Bonferroni-Holm. Of course, the minimal power of the OMT policy for $\Pi_{\theta,any}$ is much higher than that of Bonferroni-Holm.

Insight into  the reasons for the power gaps is obtained by examining the rejection regions of the different procedures (Figure \ref{fig3FWER}).
 The boundaries between one, two, or three rejections are necessarily parallel to the axes for Bonferroni-Holm but not parallel to the axes for the OMT rejection policies. Therefore, OMT policies are not monotone policies (Def. \ref{def-monotone}), and the non-monotonicity is manifest in the sloping boundaries.
For OMT policies,
rejections of hypotheses with $p$-values greater than the nominal level are possible. This is due to the structure of the optimization problem: the likelihood for three false nulls is small if $u_3$ is close to one, and therefore to maximize the objective it is preferred to reject some of the $p$-values near the diagonal rather than include rejection regions where $u_3$ is close to one (unless $u_1$ is very small), while maintaining strong FWER control.

The OMT policy for $\Pi_{\theta,any}$ rejects only the minimal $p$-value, since there is no gain in the objective function for rejecting more than one hypothesis. Interestingly, for a large range of $\theta$s the only tight constraint for the OMT policy is the global null constraint. The optimal global test statistic is $\sum_{i=1}^K\Phi^{-1}(u_i)/\sqrt{K}$. Therefore, the level $\alpha$ OMT policy for $K$ false nulls when the only tight constraint is the global  null constraint is to reject the hypothesis with minimal $p$-value if $\sum_{i=1}^K\Phi^{-1}(u_i)/\sqrt{K}<z_{\alpha}$, where $z_{\alpha}$ is the $\alpha$th quantile of the standard normal distribution.
For $K=3$, this is the OMT policy for $\theta>-0.75$ or $\theta<-1.6$, but the OMT rejection region is smaller for $\theta\in (-1.6, -0.75)$ since both the global null constraint and the constraint of FWER control when there is one false null are tight (rows 5-6 in Figure \ref{fig3FWER} ).

Interestingly, for $K=2$ the OMT policy for $\Pi_{\theta,any}$  is to the reject the hypothesis with minimal $p$-value if  $\sum_{i=1}^2\Phi^{-1}(u_i)/\sqrt{2}<z_{\alpha}$ for any $\theta<0$. This was also noted in \cite{Rosenblum14b} in a similar setting for two hypotheses.  However, for $K=3$, such a policy is no longer valid for all $\theta<0$, since the FWER when there is one false null  will be 
inflated for  $\theta\in (-1.6,-0.75)$.

\begin{table}
\caption{\label{tab-power-FWER} Average power (columns 2-4) and minimal power (columns 5-7) when all null hypotheses are false,  for different discovery policies with strong FWER control at level 0.05.   }
\begin{tabular}{|c|ccc|ccc|}
  \hline
& \multicolumn{3}{|c|}{$\Pi_{\theta,3}$}  &  \multicolumn{3}{|c|}{$\Pi_{\theta,any}$} \\
 &
Bonferroni- & OMT policy  & OMT policy & Bonferroni- & OMT policy & OMT policy\\
$\theta$ &
Holm & for $\Pi_{\theta,3}$ & for $\Pi_{\theta,any}$ & Holm & for   $\Pi_{\theta,3}$ & for  $\Pi_{\theta,any}$\\
 \hline
-0.5 & 0.0547 & 0.111  & 0.073 & 0.149 & 0.194& 0.218 \\
-1.33 & 0.241 & 0.363  & 0.247&  0.515 & 0.660& 0.742  \\
-2 & 0.530  & 0.633  & 0.323& 0.837& 0.931 & 0.968\\ \hline
\end{tabular}
\end{table}

For analysis of the OMT policy with $FDR$ control in this setting, see \S~\ref{supp:subsec-3FWER}.  An interesting finding is that when the signal is weak, the OMT policy with $FDR$ control is to reject all hypotheses if the optimal global null test is rejected at the nominal $FDR$ level (Proposition \ref{sup:prop-FDR}).

\newcommand{\vth}{\ensuremath{\vec \theta}}

\section{Beyond simple hypotheses: dealing with complex alternatives} \label{Sec:maximin}
In practical multiple testing scenarios, it is often more realistic to assume that there is no specific known alternative distribution, but that there is a family of relevant alternatives indexed by a parameter $\theta \in \Theta_A$ \citep{LehmannRomano05}. Hence, it is important to expand our results to dealing with complex alternatives.
Requiring strong control under a range of alternatives translates to requiring that the constraints in (\ref{opt}) hold for every alternative distribution (note that in multiple testing, unlike the single hypothesis case, the constraints do depend on the alternative). We limit our setting of interest to cases that have the {\em monotone likelihood ratio} property \citep{LehmannRomano05}, meaning that the p-value for each hypothesis is uniquely defined based on the cumulative distribution of the likelihood ratio under the null, regardless of the specific alternative value of the parameter. 
This property holds for large classes of distributions, including exponential families \citep{LehmannRomano05}.

We define the additional notation: assume each test  $k$ deals with a single parameter $\theta_k$, with $H_{0k}:\theta_k=0$. For a vector of potential alternatives $\vth \in \Theta_A^K$, a vector of p-values $\vu \in[0,1]^K$, and a configuration of hypotheses  $\vec h\in\{0,1\}^K$, denote the density by  $f_{\vec h,\vth}(\vec u)$, and correspondingly the error measure $Err_{\vec h,\vth}$.
 The power of the policy $\vec D$ is $\Pi_{\vth}(\vec D)$ when the parameter is $\vth$ (and the power can be any of $\Pi_{\vth,any}, \Pi_{\vth,L}$ as before). In case $\theta_1=\theta_2=\ldots=\theta_K=\theta$ we use the scalar notations $\Pi_{\theta}(\vec D), Err_{\vec h, \theta}.$

We consider two objectives:
\begin{enumerate}
    \item {\em Single objective.} Assume we have a specific alternative that is of special interest, denote it $\theta_0$, and wish to optimize the power for this selected alternative, while maintaining validity for all considered alternatives:
\begin{eqnarray}
\label{opt_comp} \max_{\vec D:[0,1]^K \rightarrow \{0,1\}^K} && \Pi_{\theta_0}(\vec D)  \\
\mbox{s.t. } &&  Err_{\vec h,\vth} (\vec D) \leq \alpha\;,\; \forall \vec h \in \{0,1\}^K,\; \vth \in \Theta_A^K. \nonumber
\end{eqnarray}
    \item {\em Maximin.} In this case, we aim  to maximize the minimal power among all  alternatives of interest $\vth \in \Theta_B^K\subseteq \Theta_A^K$ , under the same set of constraints:
\begin{eqnarray}
\label{opt_maximin} \max_{\vec D:[0,1]^K \rightarrow \{0,1\}^K} && \min_{\vth \in \Theta_B^K} \Pi_{\vth}(\vec D)  \\
\mbox{s.t. } &&  Err_{\vec h,\vth} (\vec D) \leq \alpha\;,\; \forall \vec h \in \{0,1\}^K,\; \vth \in \Theta_A^K. \nonumber
\end{eqnarray}
\end{enumerate}

Note that we do not assume that the actual parameter is the same for all alternatives, but the range of potential alternatives is the same. Hence the symmetry requirement on the resulting regions is still applicable.

These optimization problems now have, in addition to an infinite number of variables, also an infinite number of integral constraints (assuming $\Theta_A$ is an infinite set).

We are unable to offer guarantees on existence and sparsity of the optimal solutions to the problems we pose, as we have in the simple hypotheses case. Instead, we offer an approach that {\em assumes} existence of an optimal solution that can be characterized using a single value of the parameter. If the assumption holds, our proposed approach is able to find this OMT solution and --- importantly --- confirm its optimality.

Let
$\vec D^*(\theta_0,\theta)$ be the optimal solution of the optimization problem that uses fixed parameters $\theta_0$ in the objective and $\theta_A\in\Theta_A$ in the constraints:
\begin{eqnarray} \label{two_theta}
\label{opt_two} \vec D^*(\theta_0,\theta_A) = \arg\max_{\vec D:[0,1]^K \rightarrow \{0,1\}^K} && \Pi_{\theta_0}(\vec D)\\
\mbox{s.t. } &&  Err_{\vec h,\theta_A} (\vec D) \leq \alpha\;,\; \forall \vec h \in \{0,1\}^K. \nonumber
\end{eqnarray}

The following result states a sufficient condition for an optimal solution to problem \eqref{opt_comp}.

\begin{prop} \label{thm_comp}
Assume that we find a parameter value $\theta_A\in\Theta_A$ such that the solution $\vec D^*(\theta_0,\theta_A)$ controls $Err$ at level $\alpha$ at all  parameter values $\theta \in\Theta_A^K$:
$$
Err_{\vec h,\theta} \left( \vec D^*(\theta_0,\theta_A) \right) \leq \alpha\;,\forall \vec h\in\{0,1\}^K,\theta \in\Theta_A^K,
$$
then $\vec D^*(\theta_0,\theta_A)$ is the optimal solution to the complex alternative problem (\ref{opt_comp}).
\end{prop}
The following corollary simplifies the use of this result for finding $\theta_A$.
\begin{coro}\label{coro_comp}
If $\theta_A$ in the above Proposition exists, then we have:
$$
\Pi_{\theta_0} \left(\vec D^*(\theta_0, \theta_A)\right) \leq \Pi_{\theta_0} \left(\vec D^*(\theta_0, \theta)\right) \; \forall \theta \in \Theta_A.
$$
In words: the power of the optimal solution for constraints at $\theta_A$ is minimal among all optimal solutions $\vec D^*(\theta_0,\theta)$.
\end{coro}
With this corollary, we have a simple policy for trying to solve  (\ref{opt_comp}):
\begin{enumerate}
    \item Search over $\Theta_A$ to find $\theta_A = \arg\min_\theta \Pi_{\theta_0} \left(\vec D^*(\theta_0, \theta)\right)$.
    \item Check whether the control condition in Proposition \ref{thm_comp} holds.
\end{enumerate}

This approach requires solving problems of the form (\ref{two_theta}), which are instances of Problem \eqref{eq-primal}, where the parameter $\theta$ of the density function can be different in the power objective and in the constraints. It is straightforward to confirm that the results in \S~\ref{Sec:main-res} hold unchanged in this case, hence we can use the same ideas and algorithms to solve the current problems.

Next, we derive a similar sufficient condition for existence of a maximin solution, and corresponding approach for finding it.
\begin{thm} \label{thm_maximin}
Assume that we can find two values $\theta_0, \theta_A \in \Theta_A$ such that:
\begin{enumerate}
    \item  $\vec D^*(\theta_0,\theta_A)$ is the optimal solution of the single objective problem (\ref{opt_comp}) at $\theta_0$.
    \item The power of this solution at other values is higher:
    $$
     \Pi_{\theta_0} \left(\vec D^*(\theta_0,\theta_A)\right) \leq \Pi_{\vth} \left(\vec D^*(\theta_0,\theta_A)\right)\;\forall \vth \in \Theta_B^K.
    $$
\end{enumerate}
Then $\vec D^*(\theta_0,\theta_A)$ is the solution to the maximin problem (\ref{opt_maximin}). \end{thm}
The usefulness of this last result is not immediately evident, since the conditions seem harsh. As we show next, it can be practically useful when the problem is such that $\Theta_B$ has a minimal element (``closest alternative''), and there exists inherent monotonicity in the problem such that when $\theta_0$ is taken as the closest alternative, the conditions hold. 

\subsection{Example: testing  independent normal means}
Assume we are testing $K$ independent normal means with variance 1, with $\Theta_B=(-\infty, \theta_0]$ and $\Theta_A = (-\infty,0].$
We seek the maximin  rejection policy with the objective function of average power for $K=3$ false nulls, $\Pi_{\theta,K}$. By solving the optimization problem for a single $\theta<0$ constraint at a time, we compute  series of rejection policies $\vec D^*(\theta_0, \theta)$.  We identify the value of $\theta$ with minimal power, $\theta_A$,  so that $\Pi_{\theta_0,K}(\vec D^*(\theta_0, \theta_A))\leq \Pi_{\theta_0,K}(\vec D^*(\theta_0, \theta))$ for all $\theta<0$. Once we find this, we can check if
we have:
 $$\Pi_{\vec\theta,K} (\vec D^*(\theta_0, \theta_A)) \ge \Pi_{\theta_0,K} (\vec D^*(\theta_0, \theta_A))\;,\;\forall \vec \theta\in \Theta_B^K, $$ and
 $$
Err_{\vec h,\vec \theta} \left( \vec D^*(\theta_0,\theta_A) \right) \leq \alpha\;,\;\forall \vec h\in\{0,1\}^K,\vec \theta\in\Theta_A^K,
$$
 in which case by Proposition \ref{thm_maximin}, the computed solution is the maximin solution for $\Theta_B^K$. This turned out to be the case for all $\theta_0$ values in the examples considered below.

\begin{figure}[htbp]
 \begin{tabular}{cccc}
  \includegraphics[width=4cm,height=4cm,page=1]{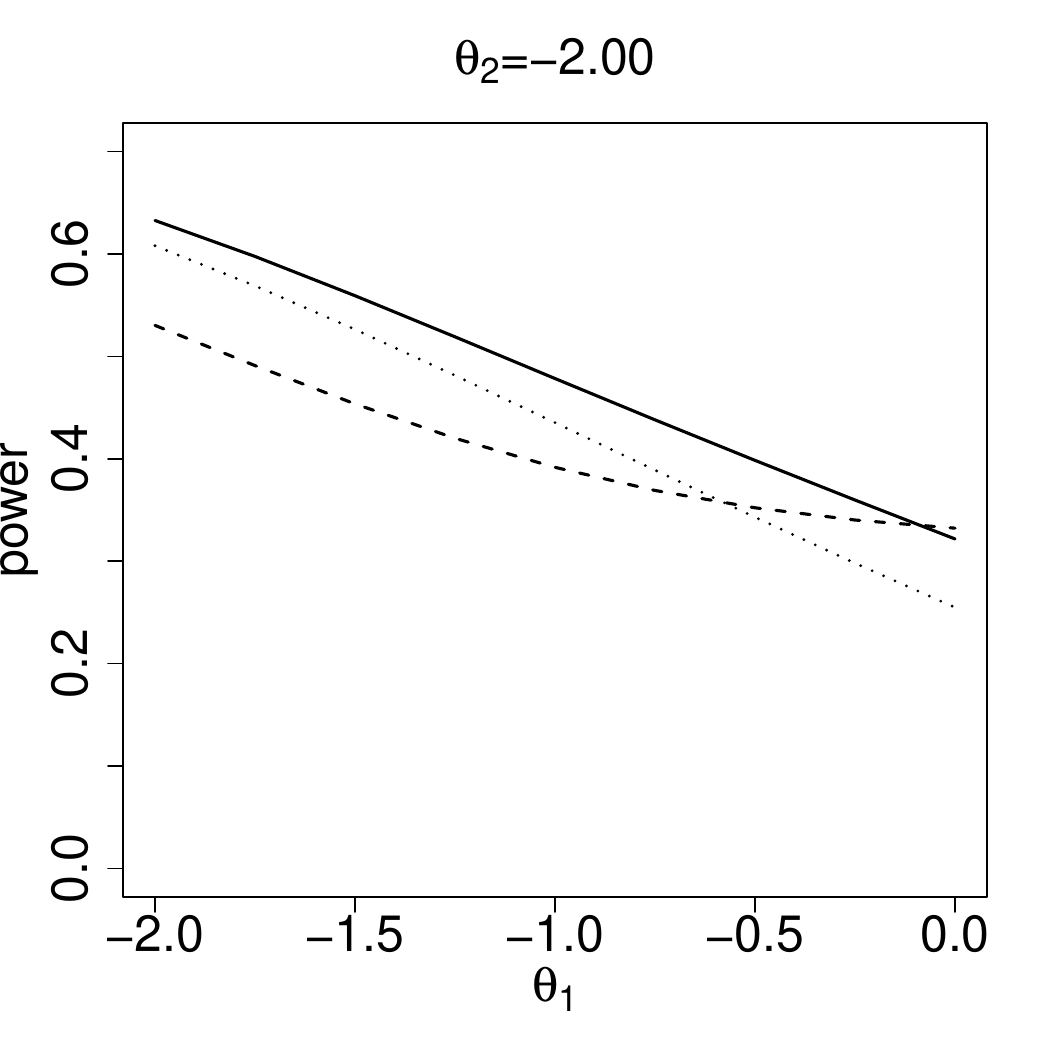} &      \includegraphics[width=4cm,height=4cm,page=2]{power-2-unequalthetas.pdf} &        \includegraphics[width=4cm,height=4cm,page=3]{power-2-unequalthetas.pdf} &    \includegraphics[width=4cm,height=4cm,page=4]{power-2-unequalthetas.pdf}
  \end{tabular}
 \caption{\label{figpowerunequalthetas} Power of the maximin (solid), closed-Stouffer (dotted) and Bonferroni-Holm (dashed) rejection policies  for normal test statistics at $\vec \theta = (\theta_1,\theta_2,-2)$. The maximin procedure is optimized for average power at $\theta_0 = -2$. }
 
\end{figure}

We examine three independent normal means with FWER control, at $\theta_0=-2$, optimized for average power. This maximin rejection policy will be used in subgroup analysis in \S~\ref{subsec-cochrane}. We examine its power at a range of $\vec \theta$ values, and compare it to Bonferroni-Holm/Sidak
, as well as to the following  closed-testing procedure \citep{Marcus76},  which is commonly applied to subgroup analysis: for each  intersection hypothesis, the sum of z-scores for the hypotheses in the intersection is the test statistic \citep{Stouffer49}.  We refer to the resulting test as closed-Stouffer.
 By definition, the maximin rejection region with strong FWER control is the most powerful at $\vec \theta =(-2,-2,-2)$. The maximin, closed-Stouffer, and Bonferroni-Holm have, respectively, an average power of 0.633, 0.609, and 0.5305 and a probability of at least one discovery  of 0.940, 0.907, and 0.837.
 Figure \ref{figpowerunequalthetas} shows the power of the three procedures for a range of $\vec \theta = (\theta_1,\theta_2, -2)$ values, where $(\theta_1, \theta_2)\in [-2,0]^2$ .  The maximin policy has better power than  closed-Stouffer for all values of $\vec \theta$: while the power gap is fairly small at $(-2,-2,-2)$, the gap increases as $\theta_1$ and $\theta_2$ approach zero. Bonferroni-Holm has a small power advantage over the maximin rejection policy at $\theta_1=0$ and $\theta_2\leq -0.5$, but it is much less powerful than maximin for smaller values of $\theta_1$ and $\theta_2$.
Overall, the maximin rejection policy has better power properties than
Bonferroni-Holm and closed-Stouffer even when one or two of the coordinates have  weaker signal than $-2$.  This suggests that in applications where the test statistics are Gaussian and independent, the maximin rejection policy is superior
 to the Bonferroni-Holm and closed-Stouffer procedures.  We describe such an application in the next section \S~\ref{subsec-cochrane}.

\begin{figure}[htbp]
  \begin{tabular}{cccc}
  $u_1=1.05e-03$ & $u_1=1.66e-02$ & $u_1=4.88e-02$ & $u_1=5.63e-02$\\
  \includegraphics[width=3cm,height=3cm]{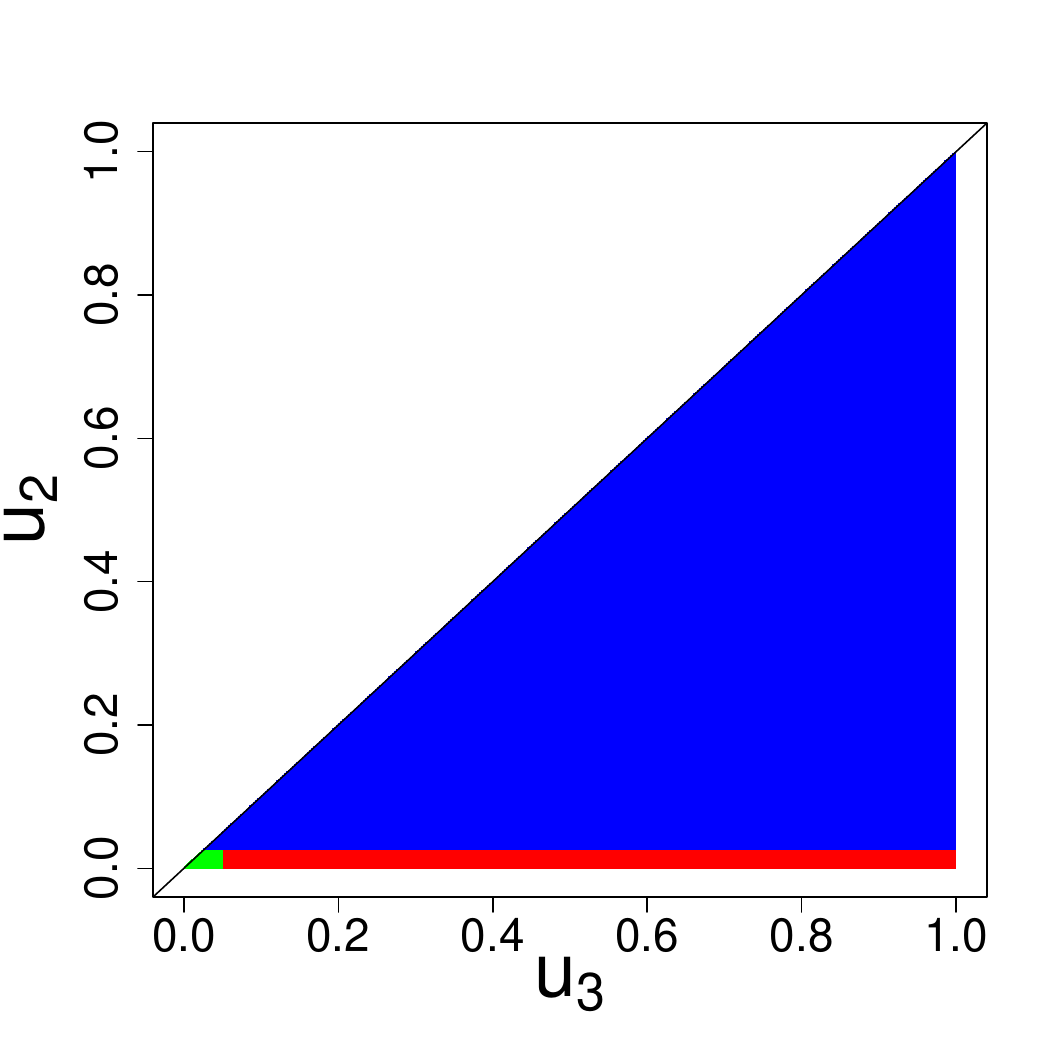} &
  \includegraphics[width=3cm,height=3cm]{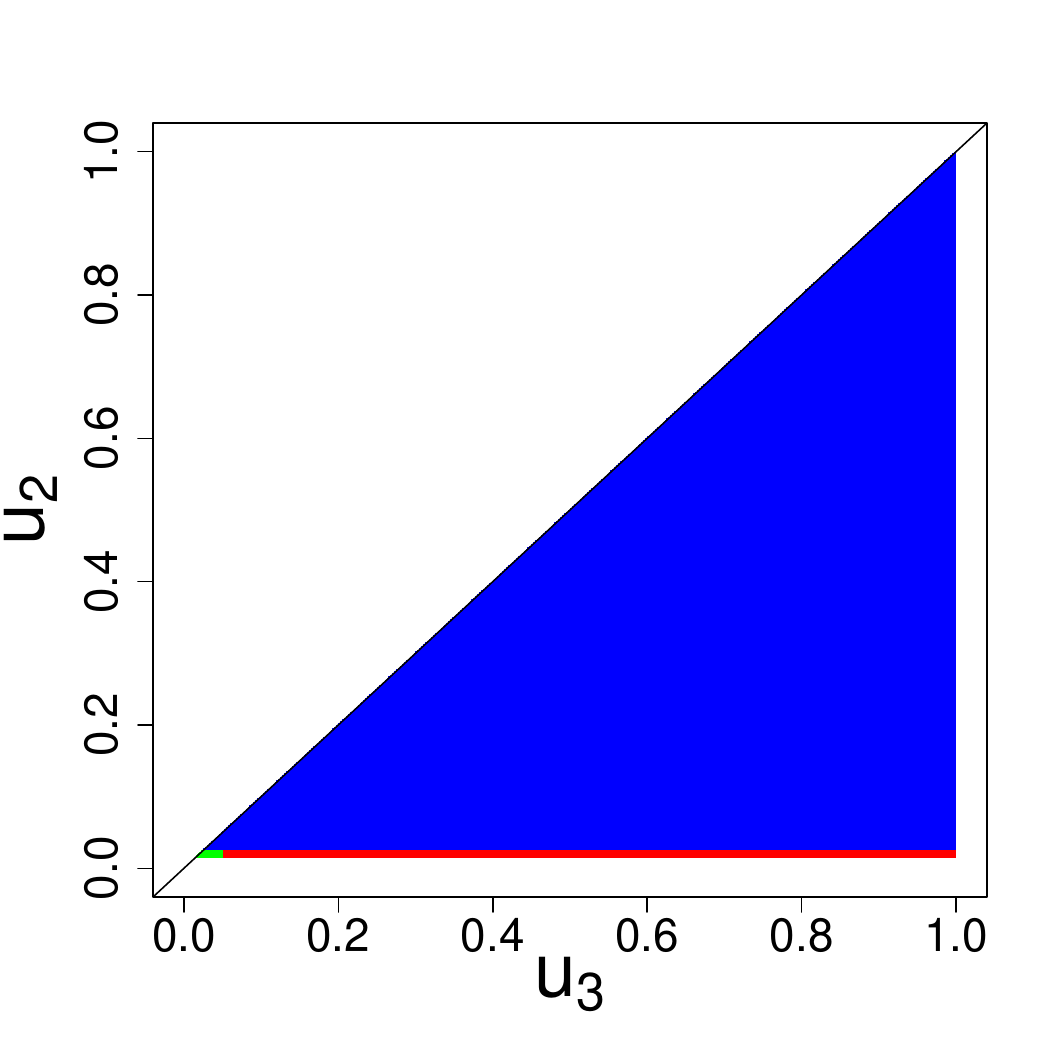} &  &
\\
\includegraphics[width=3cm,height=3cm]{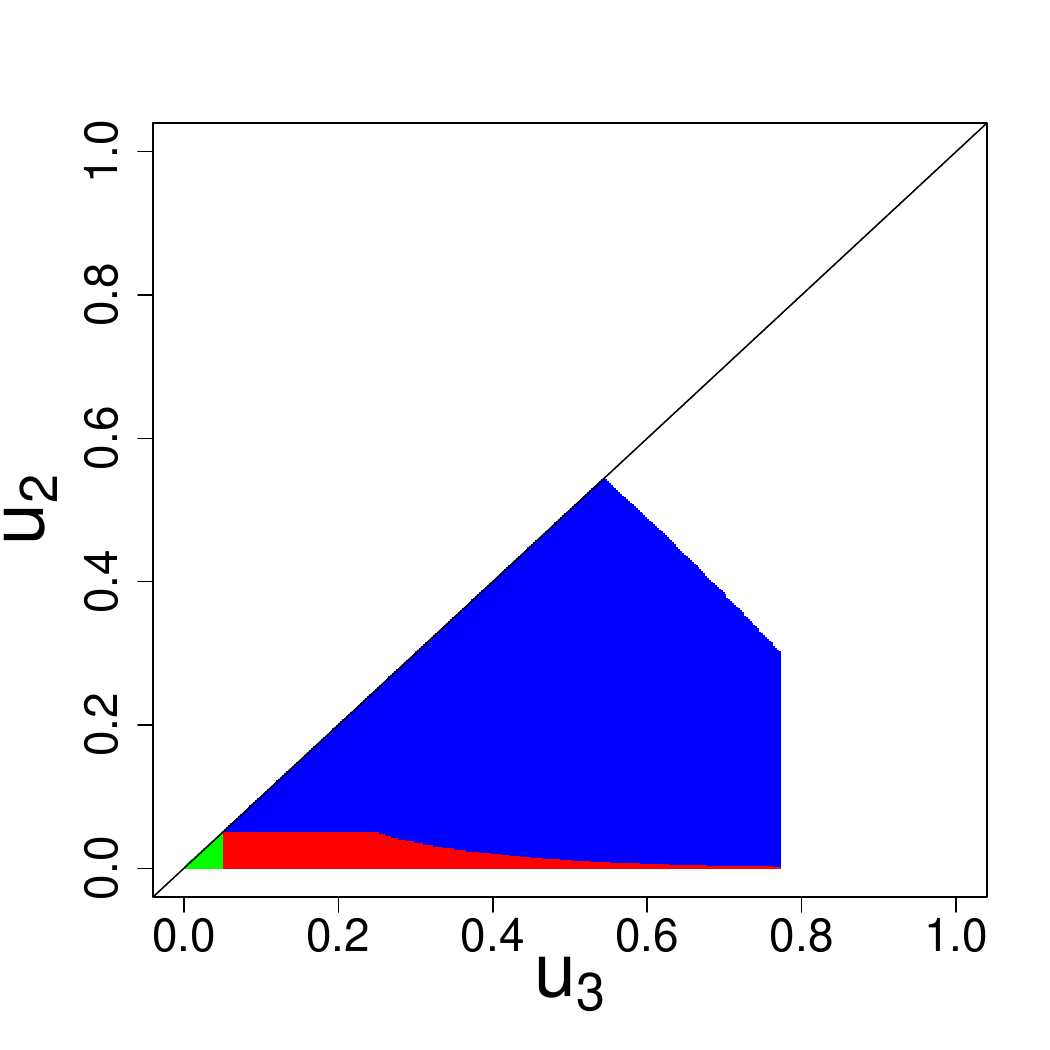}  &
\includegraphics[width=3cm,height=3cm]{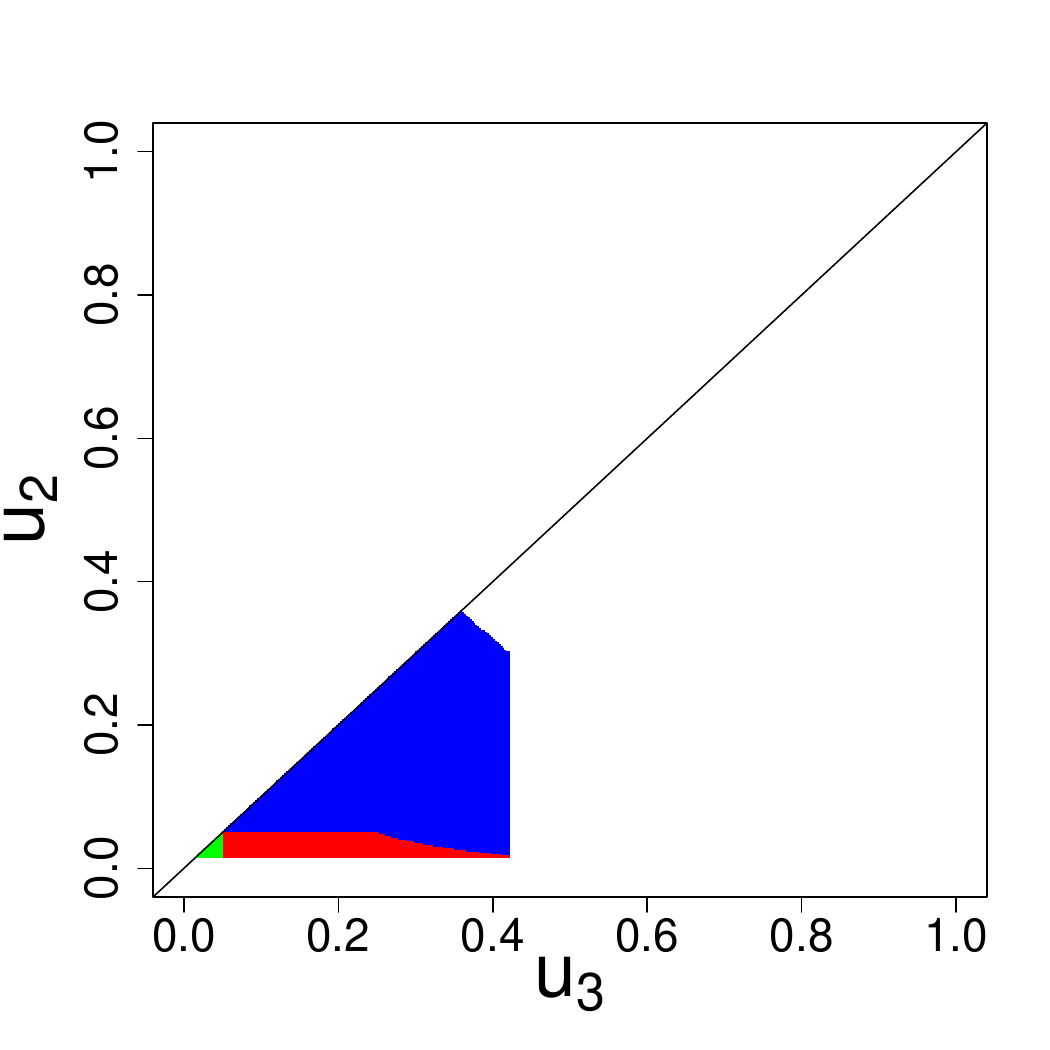}  &
\includegraphics[width=3cm,height=3cm]{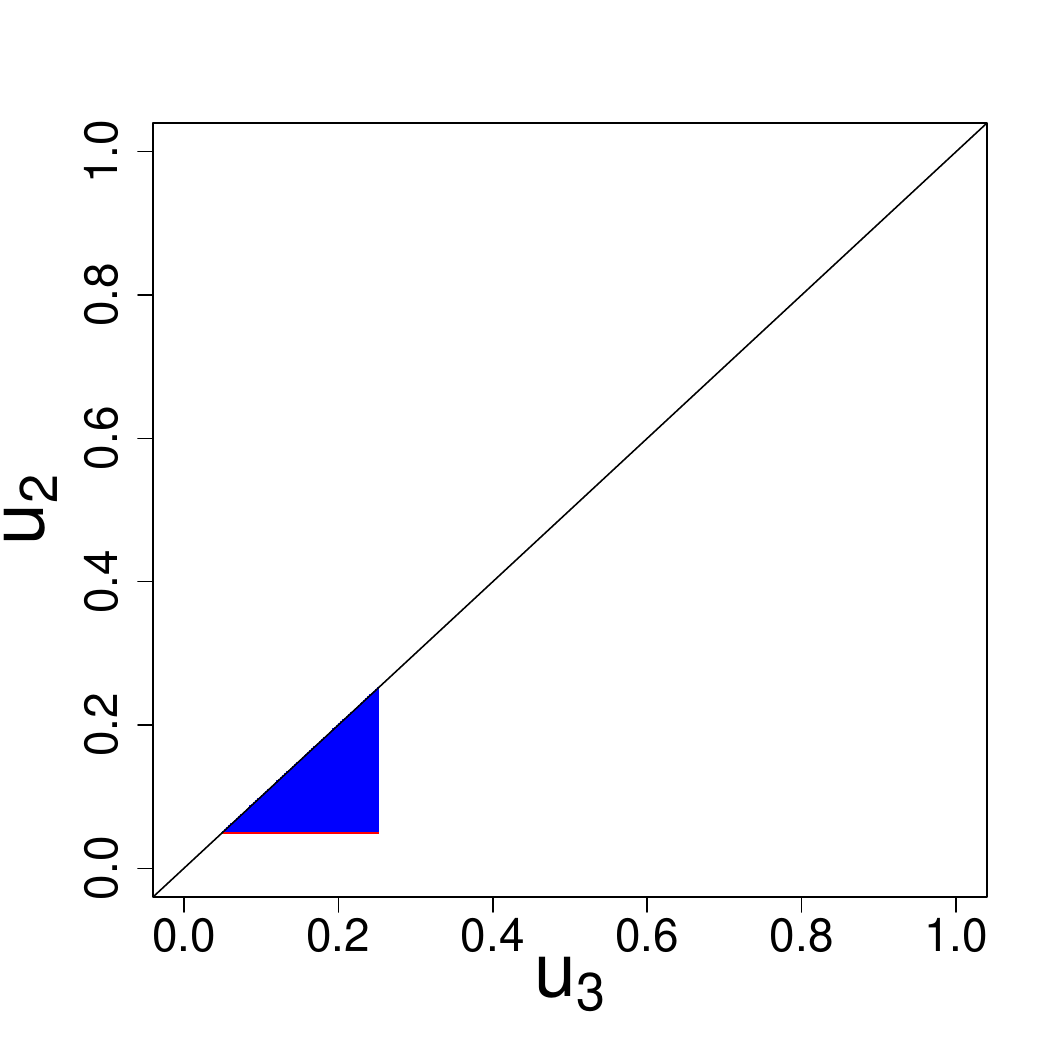}  &

\\
\includegraphics[width=3cm,height=3cm, page=58]{fwer-2-conditional-min-nomain.pdf} &
\includegraphics[width=3cm,height=3cm, page=97]{fwer-2-conditional-min-nomain.pdf} &
\includegraphics[width=3cm,height=3cm, page=113]{fwer-2-conditional-min-nomain.pdf} &
\includegraphics[width=3cm,height=3cm, page=116]{fwer-2-conditional-min-nomain.pdf}
 \end{tabular}
 \caption{ \label{fig3maximin} \small For fixed values of the minimum $p$-value, displayed are 2-dimensional slices of the following 3-dimensional rejection regions for strong FWER control at level 0.05: Bonferroni-Holm (row 1); closed-Stouffer (row 2); maximin rejection policy for optimizing the average power at $\Theta_B=(-\infty, -2]$ and $\Theta_A = (-\infty,0]$  (row 3). In green: reject all three hypotheses; in red: reject exactly two hypotheses; in blue: reject only one hypothesis. The top two right panels are empty since  Bonferroni-Holm makes no rejections if all $p$-values are greater than 0.05/3.  The last panel on the second row is empty since closed-Stouffer  makes no rejections if all $p$-values are greater than 0.05. For each panel, the rejection region is in the top right quadrant of the partition of the plane by the point $(u_1, u_1)$.}

\end{figure}

Figure \ref{fig3maximin} shows the rejection regions for the three procedures. For a 2-dimensional display, we selected slices of the 3-dimensional rejection region that are fixed by the minimum $p$-value.  We show the slices with a small minimum $p$-value, the largest minimum $p$-value for which Bonferroni-Holm still makes rejections (i.e., 0.05/3), a minimal $p$-value slightly below the nominal level, and a  minimal $p$-value above the nominal level. The boundaries between one, two, or three rejections are necessarily parallel to the axes for Bonferroni-Holm but not parallel to the axes for the closed-Stouffer or the maximin rejection policy.  Therefore, the latter two policies are not monotone policies (Def. \ref{def-monotone}), and the non-monotonicity is manifest in the sloping decision boundaries. Interestingly,  for the maximin rejection policy,
rejections of hypotheses with $p$-values greater than the nominal level are possible: if the smallest $p$-value is 0.0563, there is a fairly large region where only the smallest $p$-value is rejected, but  if the two smallest $p$-values are about the same, for a fairly large range of the maximal  $p$-value the two smallest $p$-values are rejected, and there is even a small region near the diagonal (in green color) where all three hypotheses are rejected.

\section{Application to subgroup analysis in the Cochrane library}\label{subsec-cochrane}
The Cochrane database of systematic reviews (CDSR) is the leading resource for systematic reviews in health care. Each review typically includes several outcomes, and for each outcome there may be several subgroups for a subgroup analysis. The subgroups may differ by patient characteristics (e.g., males and females), by study (e.g., studies in different geographic locations), or by intervention used \citep{Higgins11}.

We considered all the updated reviews up to 2017 in all domains. For subgroup analysis, we considered outcomes that satisfied the following criteria: the outcome was a comparison of means; the number of participants in each comparison group was more than ten; there were at least three subgroups. For simplicity, if the outcome had more than three subgroups we only considered the first three, in order to  have $K=3$ subgroup hypotheses for each outcome.   The number of outcomes that passed our selection criteria was 1321.

For each outcome, we applied the following three procedures:
the maximin procedure with $\Theta_B=(-\infty, -2]$ and $\Theta_A = (-\infty,0]$, Bonferroni-Holm/Sidak, and closed-Stouffer.  The 2-dimensional slices of their rejection policies are depicted in Figure \ref{fig3maximin}.

The lowest and highest number of discoveries are given, respectively, by
closed-Stouffer and the maximin procedure, see Table \ref{tab}.
The cross-tabulations in Table \ref{tab2} show that for every outcome in which closed-Stouffer makes discoveries, the maximin rejection policy makes discoveries as well. Moreover, in 116 outcomes, discoveries are made only in one of the two procedures Bonferroni-Holm and maximin, and in 75 of these outcomes, it is maximin that makes the discoveries.

Interestingly, there are four outcomes in which the maximin rejection policy makes exactly two discoveries, yet the other two procedures make no discoveries. The $p$-values for these four outcomes are as follows: (0.020, 0.026, 0.500); (0.033, 0.038, 0.323); (0.055, 0.055, 0.201); (0.057, 0.057, 0.500).

\begin{table}
\caption{\label{tab} Summary of discoveries made by each rejection policy, for the 1321 outcomes from the Cochrane database.}
\begin{tabular}{rrrr}
  \hline
 & maximin & Bonferroni-Holm & closed-Stouffer \\
  \hline
Average number of discoveries & 1.097 & 1.089 & 1.040 \\
Fraction with at least one discovery & 0.620 & 0.594 & 0.548 \\
   \hline
\end{tabular}
\end{table}

\begin{table}
\caption{\label{tab2} The cross-tabulation of the number of discoveries of Bonferroni-Holm and maximin and of closed-Stouffer and maximin, for the 1321 outcomes  from the Cochrane database.}
\begin{tabular}{|c|cccc|cccc|}
  \hline
 & \multicolumn{4}{|c|}{Bonferroni-Holm} & \multicolumn{4}{|c|}{Closed-Stouffer}\\
 maximin & 0 & 1 & 2 & 3 & 0 & 1 & 2 & 3  \\
  \hline
0 & 461 &  34 &   7 &   0 & 502 &   0 &   0 &   0 \\
  1 &  54 & 275 &  31 &   6 &  88 & 255 &  17 &   6 \\
  2 &  15 &  22 & 209 &  30 &   7 &  26 & 213 &  30\\
  3 &   6 &   4 &   0 & 167&   0 &   2 &   2 & 173 \\
   \hline
\end{tabular}
\end{table}

\section{Adding a weak monotonicity constraint to the OMT and maximin formulations} \label{sec:weakmon}

\subsection{Motivating examples}

\begin{figure}[htbp]
  \begin{tabular}{ccc}
  $\theta_0 = -0.5$ & $\theta_0 = -1$ & $\theta_0 = -2$\\
  \includegraphics[width=4cm,height=4cm, page=1]{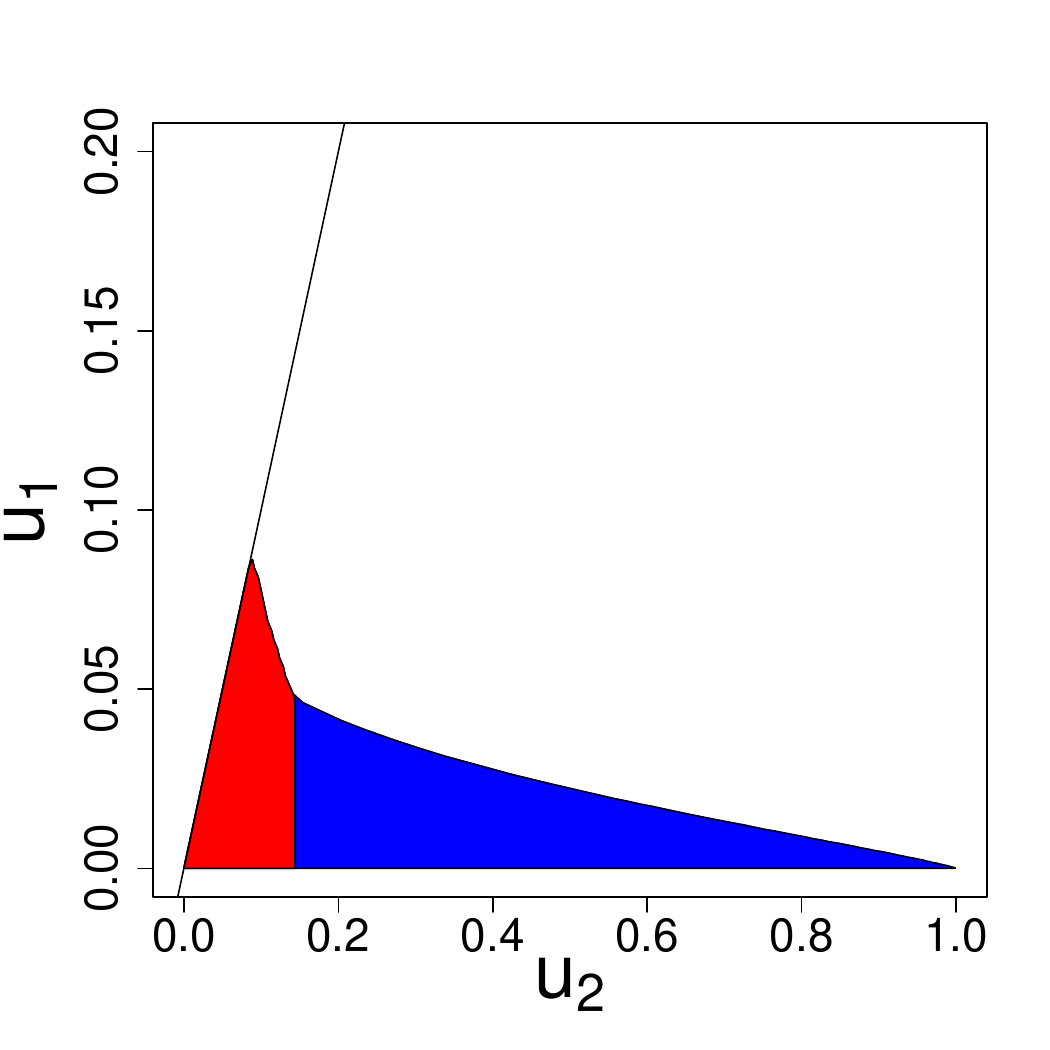} &
  \includegraphics[width=4cm,height=4cm, page=3]{maxmin.pdf} &\includegraphics[width=4cm,height=4cm, page=5]{maxmin.pdf}
  \\
  \includegraphics[width=4cm,height=4cm, page=2]{maxmin.pdf} &
  \includegraphics[width=4cm,height=4cm, page=4]{maxmin.pdf} &\includegraphics[width=4cm,height=4cm, page=6]{maxmin.pdf}
\\
  \includegraphics[width=4cm,height=4cm, page=7]{maxmin.pdf} &
  \includegraphics[width=4cm,height=4cm, page=9]{maxmin.pdf} &\includegraphics[width=4cm,height=4cm, page=11]{maxmin.pdf}
\\
  \includegraphics[width=4cm,height=4cm, page=8]{maxmin.pdf} &
  \includegraphics[width=4cm,height=4cm, page=10]{maxmin.pdf} &\includegraphics[width=4cm,height=4cm, page=12]{maxmin.pdf}
 \end{tabular}
 \caption{\label{fig2minimax} \small Rejection regions optimized for $\Pi_{\theta_0,2}$, subject to level  0.05 error control. Rows 1 and 2:  strong FWER control  with optimal and maximin procedures, respectively. Rows 3 and 4:  strong FDR control  with optimal   and maximin  procedures, respectively. In red: reject both hypotheses; in blue: reject only one hypothesis.}
 
\end{figure}

We start by examining  $K=2$ independent normal means with FWER or FDR control.
Table \ref{tab-power-maximin} shows the power comparison for various values of $\theta_0$. The maximin power is higher than that of the monotone rejection rule, but to a lesser extent than the OMT policy at a single  $\theta_0$ constraint. The power gap between the maximin and OMT solutions decreases  with $\theta_0$, and is negligible at $\theta_0=-2$. 
Figure \ref{fig2minimax} shows the optimal and maximin rejection policies. A problematic non-monotonic behavior of the maximin optimal rejection policies is manifest for both error controls: the boundary between the regions where one versus two hypotheses are rejected has a positive slope for $\theta_0\leq -1$; for $\theta_0=-0.5$ there is a gap between the regions where two hypotheses versus one hypothesis are rejected. These regions contradict the reasonable principle that if $(u_1,u_2)<(u_1', u_2')$ elementwise,  and the policy at $(u_1', u_2')$ is rejection of both hypotheses, then at $(u_1, u_2)$  both hypotheses should also be rejected. As the signal strength of the objective function $|\theta_0|$ increases, this undesired behaviour is less pronounced, as manifested by the slope on the boundary between red and blue regions being steeper.
The counter-intuitive rejection regions are due to the fact that it is possible to add pieces to the rejection region without violating the error control. For example, the maximin optimal rejection policy for $\theta_0 = -0.5$ occurs at $\theta_A = -1.29$ for strong FWER control and at $\theta_A = -1.36$ for strong FDR control. The chance of both $p$-values being of similar value and not very small is negligible if exactly one hypothesis is at $\theta_A$, hence the penalty for the red regions at $\theta_A$ is negligible, but the added power at $\theta_0\gg \theta_A$ is non-negligible. For a discussion of the advantages and disadvantages of such counter-intuitive rejection regions, see \cite{Perlman99}.

\begin{table}
\caption{\label{tab-power-maximin} Average power, $\Pi_{\theta_0,2}$, for the following rejection policies. For strong FWER control: Bonferroni-Holm (column 2); OMT policy for $\Pi_{\theta_0,2}$  (column 3); maximin optimal for $\Pi_{\theta,2}$ with $\Theta_B = \{\theta: \theta\leq \theta_0 \}$ and $\Theta_A = \{\theta: \theta\leq 0 \}$  (column 4). For strong FDR control: MABH (column 5); OMT policy for $\Pi_{\theta_0,2}$  (column 6); maximin optimal for $\Pi_{\theta,2}$ with $\Theta_B = \{\theta: \theta\leq \theta_0 \}$ and $\Theta_A = \{\theta: \theta\leq 0 \}$  (column 7).}
\begin{tabular}{|c|ccc|ccc|}
  \hline
  & \multicolumn{3}{|c|}{Strong FWER control} & \multicolumn{3}{|c|}{Strong FDR control}  \\ \hline
$\theta_0$ & Bonferroni-Holm & OMT   &  maximin & MABH  &  OMT  &  maximin \\ \hline
-0.5 & 0.076 & 0.118& 0.099 & 0.086 & 0.174 & 0.129\\
-1&0.184&0.251&0.237 & 0.214 & 0.326 &  0.296\\
-2 & 0.581 &0.637&0.636& 0.660&0.734& 0.733\\ \hline
\end{tabular}
\end{table}

\subsection{Enforcing monotonicity}
A major concern with the optimal solutions in Figure \ref{fig2minimax} is the lack of {\em weak monotonicity}  in the resulting region. Formally we define weak monotonicty as requiring that if $\vu \preceq \vv$ then $\vD^*(\vu) \succeq \vD^*(\vv)$, where $\preceq, \succeq$ are the Euclidean partial order relations (that is, they hold iff simple inequality holds for every coordinate). 

In this section we reformulate our previous problems with the additional weak monotonicity requirement, and demonstrate that essentially all previous results hold, and we can therefore find {\em mono-OMT} and {\em mono-maximin} solutions similarly. We maintain Assumptions \ref{ass:exch} and \ref{ass:monoLR} made in the OMT policy derivation. 

Rewriting our generic problem formulation with  LR-ordering and symmetry, and the additional monotonicity constraint we obtain: 
\begin{eqnarray}
\label{opt-mon} \max_{\vec D:Q \rightarrow \mathcal D} && \Pi(\vec D)\\
\mbox{s.t. } &&  Err_{\vec h_L} (\vec D) \leq \alpha\;,\; 0\leq L < K \nonumber \\
&&  \vD(\vu)\succeq \vD(\vv) \;,\; \forall  \vu \preceq \vv. \nonumber
\end{eqnarray}

We integrate the monotonicity constraint into the domain by changing the domain of the function $\vD,$ such that instead of considering all functions $\vD \in {\mathcal D}^Q,$ we now limit ourselves to functions that also comply with the monotonicity constraint for all relevant pairs. Writing it like that, the Lagrangian is unchanged from Eq.~(\ref{eq:Lagrangian}), but we have to make sure we only consider  legal (monotonic) decision rules $\vD$.

We find the solution by the following steps: 
\begin{enumerate}
    \item Design a policy to search the space of integral constraint Lagrange multipliers $\vmu = \mu_0,\ldots,\mu_{K-1}$
    \item Given $\vmu,$ solve the following isotonic regression problem:  
    \begin{eqnarray}\label{opt-mono-simple}
\max_{\vec D:Q \rightarrow \mathcal D} && L(\vec D, \mu)\\
\mbox{s.t. } 
&&  \vD(\vu)\succeq \vD(\vv) \;,\; \forall  \vu \preceq \vv. \nonumber
\end{eqnarray}
    \item Search for $\vmu^*$ that attains primal feasibility as in the regular OMT case (i.e., complies with the conditions in Equation \eqref{eq:mucond}). 
\end{enumerate}

The main increased complexity of this solution of weakly-monotone OMT relative to the simple OMT case is that step 2 above requires finding a solution to an infinite linear isotonic regression problem, compared to the simpler characterization in Equations \eqref{Dmu1}--\eqref{Dmu2}. 
We approximate the solution by solving finite linear isotonic regression problems, for which efficient algorithms exist (\citealp[and references therein]{Luss10,Luss12}). 

The results in \S~\ref{Sec:maximin} do not depend on the form of the monotonicity constraints in the problem, and hence all these results persist as-is for the weakly-monotone case. In particular, Theorem~\ref{thm_maximin} carries through unchanged and gives an approach for finding maximin solutions under weak monotonicity, if its sufficient conditions hold. 

\subsection{Illustration} \label{sec:egweakmon}
Two important questions that arise about the weakly-monotone version of OMT and maximin is whether we can solve it in practice given the increased complexity of solving many isotonic regression problems, and if we can, what is the power loss from imposing the weak monotonicity requirement? In particular, how much of the power advantage over standard competitors is lost? 

To address these questions and illustrate solving weakly-monotone OMT and maximin problems, we apply the methodology to two clearly non-monotone examples in Figure~\ref{fig2minimax}: 
\begin{enumerate}
    \item Maximin solution for FWER control, $\Theta_B  =  (-\infty, -1)$ and $\Theta_A = (-\infty,0)$ (middle plot of second row in the figure).
    \item OMT solution for FDR control, $\theta=-1$ (middle plot of third row). 
\end{enumerate}

\begin{figure}[htbp]
  \begin{tabular}{cc}
   {\Large Optimal} & {\Large Weakly monotone}\\
  \includegraphics[width=5cm,height=5cm, page=2]{FWER-weaklymonotone.pdf} &\includegraphics[width=5cm,height=5cm,page=1]{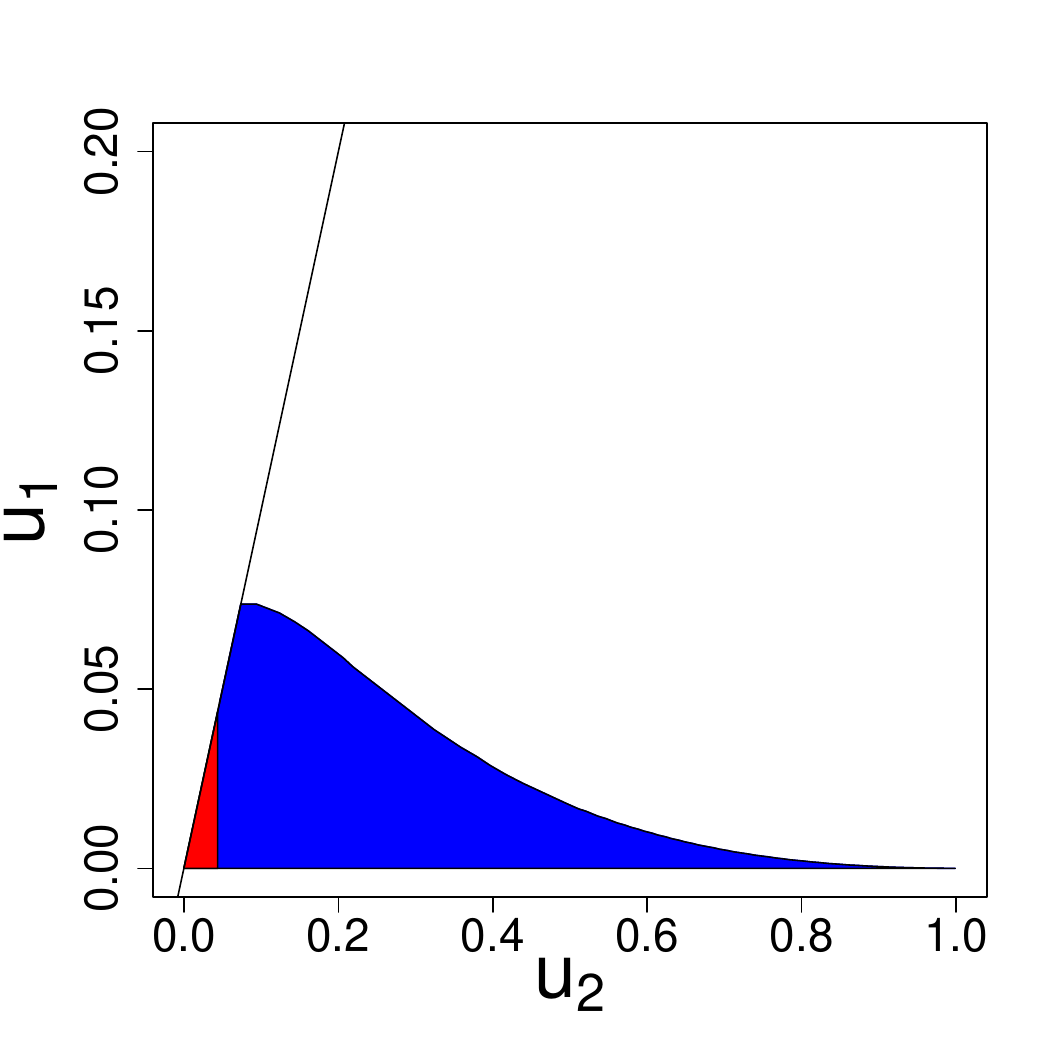}
\\
  \includegraphics[width=5cm,height=5cm, page=2]{FDR-weaklymonotone.pdf} &\includegraphics[width=5cm,height=5cm,page=1]{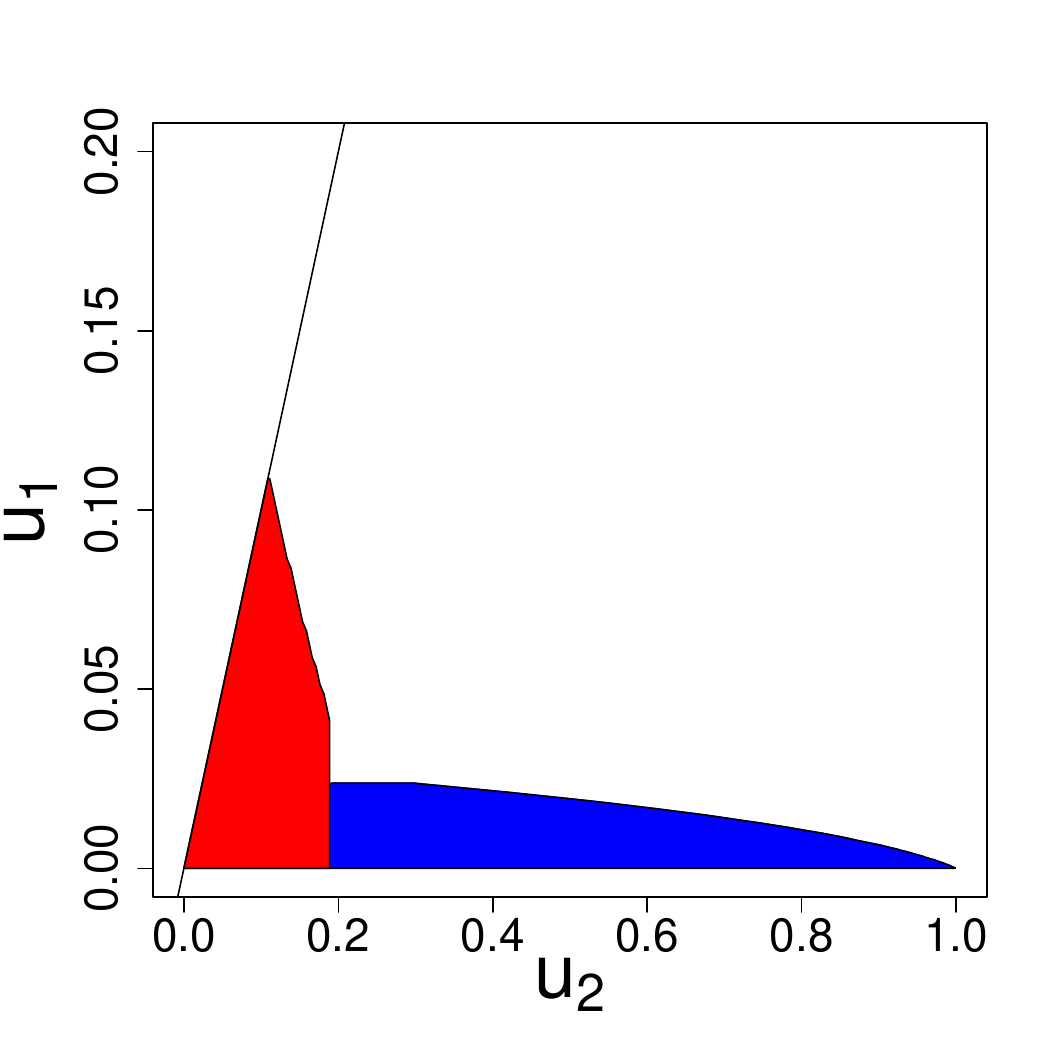}
 \end{tabular}
 \caption{\label{fig-weakmon} \small Comparing previously shown non-monontone optimal rejection regions (left column, See Figure \ref{fig2minimax} and Table \ref{tab-power-maximin}) to weakly-monotone optimal rejection regions for the same problems. Top row: maximin for FWER control with $\Theta_B=(-\infty,-1]$ and $\Theta_A=(-\infty,0)$. Bottom row: OMT for FDR control with $\theta=-1$. The power loss is minimal: from $0.237$ to $0.231$ in the first row, and from $0.326$ to $0.325$ in the second --- see Table \ref{tab-power-maximin} for comparison to existing alternatives.}
 
\end{figure}

In Figure \ref{fig-weakmon} we show the resulting weakly-monotone rejection regions (right column) and compare them to the non-monotone regions from   Figure~\ref{fig2minimax} (left column). In terms of power loss, the effect of enforcing monotonicity is quite minor in these examples: comparing to the results in Table \ref{tab-power-maximin}, we see a decrease from $0.237$ to $0.231$ in the minimal power of the maximin FWER procedure, attained at $\theta=\theta_0=-1$, still much higher than Bonferroni-Holm (and also of closed-Stouffer, whose power is 0.207). For the OMT FDR procedure, we see a very minimal decrease in power from $0.326$ to $0.325$, compared to MABH which only attains $0.214$. 

\section{Related work on optimal and maximin procedures}\label{Sec:related work}
\subsection{Previous work on optimal FWER control}
The  simplest class of FWER controlling procedures is that  of single-step procedures, where the decision whether to reject a hypothesis is only based on the test statistic (or p-value) for that hypothesis. For the weighted Bonferroni procedures,  weights to maximize the average power  have been considered, e.g., in  \cite{spjotvoll72, Westfall98, Dobriban15}. 

Optimality results are also available for a  more general class of FWER controlling procedures, which requires the selection rules to be monotone.
\begin{definition}[\citealp{Lehmann05}]\label{def-monotone}
A decision rule $\vec D:[0,1]^K\rightarrow \{0,1 \}^K$  is said to be monotone if $u'_i\leq u_i $ for $D_i(\vec u)=1$ but $u_i'>u_i$ for $D_i(\vec u)=0$ implies that $\vec D(\vec u)=\vec D(\vec u')$. \\
In words, if the value of the rejected $p$-values is decreased, and the value of the non-rejected $p$-values is increased, the set of rejections remains unchanged.
\end{definition}
If restricted to monotone decision rules,  the optimal procedure is in the family of stepwise procedures \citep{Lehmann05} \footnote{The decision on whether to reject in stepwise procedures depends on the rank of the $p$-value: step-down procedures begin by looking at whether the most significant $p$-value should be rejected; step-up procedures begin by looking at the least significant $p$-value .}.

The restriction to monotone decision rules excludes closed testing procedures \citep{Marcus76} that are based on  combined test statistics (e.g., based on the sum of the $z$-scores) for testing the intersection hypotheses.
Such procedures  have been shown to have better power than stepwise procedures, unless there is a single strong signal among a group of otherwise null or very weak signals  (in which case step-down tests are best), see e.g., \cite{Lehmacher91,Bittman09}.

A direction that is most similar to ours, of pursuing optimal power with strong FWER control, with no restriction on the form of regions generated, was explored in \cite{Rosenblum14}, and optimal rejection regions were presented for $K=2$ which are clearly not in the family of monotone selection rules. While the derivation of optimal monotone selection rules in \cite{Lehmann05} is relatively easy, the derivation of the optimal rules as suggested in \cite{Rosenblum14} is computationally difficult. Their optimization technique is based on discrete approximation of the relevant probabilities, which is computationally feasible with two hypotheses, but may be infeasible for more hypotheses. They leave the extension to more than two hypotheses for future research.

Our objective of maximizing power with strong FWER control is similar to that of  \cite{Rosenblum14}. However, we address the optimization of the continuous problem in a general framework, which is  different from their approach. From the equations of the optimal solution, we demonstrate how we can gain insight into the nature of the rejection region.  We demonstrate for $K=2,3$ hypotheses the significantly higher power that can be obtained over the stepwise procedures of \cite{Lehmann05}, and we show that the optimal rejection regions are not monotone.


\subsection{Previous work on optimal FDR control}
The best known FDR controlling procedure is the Benjamini-Hochberg (BH) procedure \citep{Benjamini95}, which has been shown to perform nearly optimally for various loss functions assuming the hypotheses are exchangeable, when the fraction of null hypotheses is close to one \citep{Genovese02}.  Asymptotically, as the number of hypotheses grows to infinity, \cite{Castro17} showed that  the BH procedure is optimal in some sense; \cite{Finner09} derived an asymptotically optimal rejection curve under some restrictions on the possible rejections.

An important line of work in recent years concerns control of the marginal FDR, $\mE(V)/\mE(R)$, where the expectations are over $\vec h$ and $\vec u$, assuming the Bayesian setting that the hypotheses come from the two-group model.
\citep{Efron08,Sun07,Storey07}. 

Our objective of maximizing power with strong FDR control stands apart from  this line of work in an important way: our FDR control guarantee is for the realized $\vec h$, and it is non-asymptotic. Moreover,
we do not assume knowledge of the percentage of true null hypotheses.
We present the optimization of the continuous problem, and show from the equations of the optimal solution how we can gain insight into the nature of the rejection region.  We demonstrate for $K=2,3$ hypotheses the significantly higher power that we can obtain over the BH procedure as well as over the procedure of \cite{Solari17} which provides a small but uniform improvement over the BH procedure. We further show that the optimal rejection region is not monotone, but that monotonicity constraints can be enforced in order to receive a weakly-monotone OMT procedure (see \S~\ref{sec:weakmon} for details).

\subsection{Previous work on optimal maximin procedures} We are not aware of much related work. We note a line of work under the name Generalized Neyman Pearson (GNP), which deals with finding minimax tests for single hypotheses when  the null and alternative are allowed to be complex. In this setting, the simple NP solution no longer holds, but extensions using convexity and duality arguments allow asserting the existence of minimax-optimal solutions in certain cases \citep{Cvitanic01,Rudolff10}. Our maximin OMT problems are more complex because of the structure and number of constraints, and our policy in \S~\ref{Sec:maximin} of deriving testable sufficient conditions rather than theoretical guarantees reflects that.

\renewcommand{\vv}{\ensuremath{\vec v}}

\newcommand{\CJ}{\ensuremath{{\cal  J}}}
\newcommand{\CL}{\ensuremath{{\cal  L}}}
\newcommand{\CU}{\ensuremath{{\cal  U}}}

\section{Discussion and conclusion}\label{Sec:discuss}

We present a complete mathematical treatment of OMT procedures for multiple testing of exchangeable simple hypotheses, with demonstration of the resulting solutions for $K=3$ hypotheses and their power advantage over existing alternatives. In \S~\ref{Sec:maximin} we expand the results to the case of complex alternatives, which is the relevant setup in most practical applications, and offer sufficient conditions for maximin solution, which control false discovery at all alternatives, while maximizing the minimal power for the range of relevant ones. We emphasize that this maximin solution does not require that all alternatives are the same, and in that sense it also relaxes the exhangeability requirement from the simple hypotheses solution.
Critically, we demonstrate that these sufficient conditions hold for testing independent normal means, and in \S~\ref{subsec-cochrane} use this  for subgroup analysis, generating more discoveries than Bonferroni-Holm/Sidak and closed-Stouffer's approaches.

Our development has focused on establishing solvability of OMT problems, and solving relatively low dimensional instances numerically, up to $K=3$. In some modern applications, $K$ can be in hundreds, thousands or even millions (like in Genome Wide Association Studies). To address feasibility of solution for larger $K$, we need to consider the computational complexity of numerical solution, and in particular its dependence on $K$. There are three components to the computation:
\begin{enumerate}
    \item Searching in parameter space for the $K$ Lagrange multipliers which solve the problem.
    \item For each set of multipliers considered, performing numerical integration over the set $Q$ in the $K$-dimensional hypercube.
    \item For each evaluation of the integrand in the integration, calculating the coefficients in (\ref{FWER-const},\ref{FDR-const}).
\end{enumerate}
The complexity of the first two items depends on the specific algorithms used for search and integration, of which there is a large variety \citep{Press07}, and identifying the best approaches for our type of problems is a topic for future research. For the third item --- calculation of coefficients for the linear constraints --- we can make some progress. The representation in (\ref{FWER-const},\ref{FDR-const}) appears to be exponential in $K$, however it is easy to see that these coefficients can be calculated in complexity $O(K^2)$ using a dynamic programming approach, for independent hypotheses (details in \S~\ref{supp:dyn}). Hence by combining state of the art approaches for parameter optimization and numerical integration, with efficient calculation of the coefficients at each integration point, problems of dimension  higher than $K=3$ can be solved exactly and efficiently. It seems quite clear, however, that to go to dimensions in the thousands or higher, approximations would be required.  One direction for such approximations is the use of hierarchical controlling procedures, where hypotheses are divided to groups, within each group an optimal testing procedure is employed, and the results are summed up using group-aggregation techniques.
For example, for an MTP with $N\times K$ hypotheses,  if we have optimal rejection policies for $K$ hypotheses, we can adjust the level of testing within each of  $N$ groups of $K$ hypotheses in order to solve the bigger problem with the same error guarantee. Specifically, for FWER control at level $\alpha$, we can apply the optimal rejection policy at level $\alpha/N$ for each group of $K$ hypotheses, and this procedure will clearly be far more powerful than the Bonferroni procedure on the  $N\times K$ hypotheses $p$-values. For FDR control, the level of the optimal rejection policy within each group of $K$ hypotheses may be  closer to $\alpha$ than to $\alpha/N$ \citep{Efron08, BB14}. The gain over the BH procedure on all   $N\times K$ hypotheses may be substantial.

As far as we know, this is the first work that shows that   the objective and constraints are linear in the decision function, thus enabling the computation of   optimal rejection regions for FDR control.  Similar steps can be followed to establish that other objectives that are of interest with FDR control, such as 
expected weighted loss minimization \citep{Sun07}, are also linear in the decision function.  
If in  addition the number of (or a lower bound on) true nulls $K-L$ out of the $K$ hypotheses is known (or can be estimated), then this knowledge can further be exploited to define an easier optimization problem with at most $L$ constraints. Such a procedure will naturally have higher power than the optimal procedure with $K$ constraints, and it can be interesting to examine its power gain over  adaptive methods for FDR control available in the literature, e.g., \cite{Storey03,Benjamini06}. For large scale inference problems (typically, $K$ at least few thousands), the two-group model, first introduced by  \cite{Efron01}, is often assumed. Using this model, inference is typically based on methods that control the mFDR \citep{Sun07, Efron10}. Our infinite-dimensional formulation of the hypothesis testing problem led us  to find the OMT policy with  FDR and positive FDR \citep{Storey03} control within the two-group model framework in a companion paper 
\citep{Heller19}. By assuming the two-group model, the number of error constraints is reduced to one, since the error is the expectation over all possible configurations $\vec h$. For example, for FDR control the single constraint is $\mE_{\vec h, \vec u} \left(  \frac{V}{R}; R>0 \right) \leq \alpha$.  Therefore, with computational shortcuts, we were able to find the OMT policies for practically any $K$.

For FWER control, hierarchical procedures that base the decision for a hypothesis on the values of all test statistics have been advocated on an intuitive basis for the setting of non-sparse signals, but without the justification by an optimality theory \citep{Marcus76, Lehmacher91}.
\cite{Lehmann05} developed an optimality theory  for procedures restricted to be monotone (which exclude   procedures suggested in \cite{Marcus76, Lehmacher91}). Under this requirement, optimal testing procedures can take simple forms like being limited to ``step-down'' rules \citep{Lehmann05}, and these can be derived relatively easily, with no need for complex methodology we develop here. However, the OMT regions we obtain for all problems we consider are not monotone according to this very strict definition. The reward is significantly higher power than step-down procedures can supply for such problems (Table \ref{tab-power-FWER}).

We may still ask what constitutes a ``reasonable'' test, and what happens when optimal tests do not comply with reasonableness expectations. This issue has been raised in other contexts in \cite{Perlman99}.  The  form of monotonicity that appears to be uniformly desired is what we term ``weak monotonicity'' in \S~\ref{sec:weakmon}, implying that smaller  $p$-values cannot result in  less rejections. Surprisingly, some of our derived optimal procedures do not comply with this seemingly sensible requirement, including FDR regions in Figure \ref{fig3FDR}, and more pronounced, maximin regions in Figure \ref{fig2minimax}.  To accommodate this requirement, we offer a formulation which adds it to the optimization problem, show that it can be solved, but at some cost, both in terms of computational complexity, and power. In the examples we tested 
, the power loss is minimal, but more research is required to understand whether it can be more substantial in other cases.  


We leave for future research examination of the utility OMT policies  for other power functions, such as linear combinations of the functions we propose here, which can also be solved within our framework. We also leave for future research consideration of other error measures (see Table 1 in \cite{Benjamini10} for a list of measures which can be viewed as generalization of the FWER and FDR). 

\section*{Acknowledgements}
The authors thank  Yoav Benjamini, Marina Bogomolov, Thorsten Dickhaus, Jelle Goeman, Abba Krieger, Aaditya Ramdas, Ryan Tibshirani, and Wenguang Sun for useful discussions and comments on a previous version of the manuscript.  The authors also thank Ian Pan and Orestis Panagiotou for providing the processed CDSR data.
This research was supported by Israeli Science Foundation grant 2180/20.

\appendix 

\section{Proofs}\label{supp:proofs}
\subsection*{Proof of Theorem \ref{lemma-weakmonotone}}

Given a candidate symmetric solution $\vec D$, we prove the theorem by constructing an alternative solution $\vE$ that complies with the condition and has no lower objective and no higher constraints than $\vec D$. For every pair of indexes $1\leq i<j\leq K$, define:
$$ A_{ij} =\left\{\vec X :  \Lambda_i> \Lambda_j,\;  D_i(\vec X) =0, D_j(\vec X)=1 \right\}$$.

By the symmetry requirement of $\vec D$, there is a symmetric set
$$ A_{ji} =\left\{ \vec X :  \Lambda_j> \Lambda_i,\;  D_j(\vec X) =0, D_i(\vec X)=1 \right\},$$
where
$\vx \in A_{ij} \Leftrightarrow \sigma_{ij}(\vx) \in A_{ji},$
and $\sigma_{ij}$ is the permutation that switches the data for hypotheses $i,j$ only. 

We will now examine the solution $\vE$ which is equal to $\vec D$ everywhere, except on the sets $A_{ij}, A_{ji}$, where it switches the value of coordinates $i,j$:
$$
E_k(\vec X) = \left\{ \begin{array}{ll} D_k(\vec X) & \mbox{if } \vec X \notin A_{ij}\cup A_{ji} \mbox{ or } k\notin \{i,j\}\\
D_j(\vec X) & \mbox{if } \vec X \in A_{ij}\cup A_{ji} \mbox{ and } k=i \\
D_i(\vec X) & \mbox{if } \vec X \in A_{ij}\cup A_{ji} \mbox{ and } k=j
\end{array} \right. .
$$
We now show the following:
\begin{enumerate}
    \item For all power functions we consider, $\Pi(\vE) \geq \Pi (\vec D)$.
    \item For all constraints we consider $Err_{\vec h}(\vE) \leq  Err_{\vec h}(\vec D)$.
\end{enumerate}
Therefore $\vE$ is an improved solution compared to $\vec D$. This can be done for all $i,j$ pairs, and we end up with $\vE$ which has the desired property and is superior to $\vec D$. This holds in particular for $\vec D=\vec D^*$ the optimal solution, hence we are guaranteed to have a weakly monotone optimal solution in the likelihood ratio statistics.

It remains to prove properties 1,2 above. For the power, notice that 
 we can write Eqs. (\ref{pow1},\ref{pow2}) as:
$$\Pi_L(\vec D) =  \int_{\Re^K}  r\left(\vec D(\vec X)\right) \mathcal L_{\vec h_L}(\vec X) d\vec X,$$
where $r$ is {\em arrangement increasing} as a function of $\vec D(\vec X)$, that is: if we permute the coordinates of $\vec D(\vec X)$ such that the first $L$ associated with non-nulls increase, and the last $K-L$ associated with nulls decrease, then $r(\vec D(\vec X))$ increases.
Considering a specific pair of indexes $i<j$ above, if $i>L$ (both nulls) or $j\leq L$ (both non-nulls), then by symmetry of $\vec D$ and $
\vh$-exchangeability we have that $\Pi_L(\vec D) = \Pi_L(\vE).$ If $i\leq L$ but $j>L,$ we have that $\forall \vec X \in A_{ij}$:
\begin{eqnarray*}
&&r\left(\vec D(\vec X)\right) = r\left(\vE\left(\sigma_{ij}(\vec X)\right)\right)\\
&&r\left(\vec D\left(\sigma_{ij}(\vec X)\right)\right) = r \left(\vE(\vec X)\right)\\
&&r\left(\vec D(\vec X)\right) \leq r \left(\vec D\left(\sigma_{ij}(\vec X)\right)\right),
\end{eqnarray*}
where the last relation is the arrangement increasing property. We also have $$\forall \vec X \in A_{ij},\; \mathcal L_{\vec h_L}(\vec X) \geq \mathcal L_{\vec h_L} \left(\sigma_{ij}(\vec X)\right),$$ because of Assumption \ref{ass:monoLR} ($h_i=1$, $h_j=0$, and $\Lambda(\vx_i)>\Lambda(\vx_j)$ for $\vx\in A_{ij}$).
Hence we get:
\begin{eqnarray*}
&\int_{A_{ij}\cup A_{ji}}& r\left(\vec D(\vec X)\right) \mathcal L_{\vec h_L} (\vec X)\;  d\vec X  = \int_{A_{ij}}  r\left(\vec D(\vec X)\right) \mathcal L_{\vec h_L} (\vec X) + r\left(\vec D\left(\sigma_{ij}(\vec X)\right)\right) \mathcal L_{\vec h_L}\left(\sigma_{ij}(\vec X)\right)\; d\vec X \stackrel{(*)}{\leq} \\
&& \int_{A_{ij}} r\left(\vec D\left(\sigma_{ij}(\vec X)\right)\right) \mathcal L_{\vec h_L} (\vec X) + r\left(\vec D(\vec X)\right) \mathcal L_{\vec h_L}\left(\sigma_{ij}(\vec X)\right)\; d\vec X  =\\ &&\int_{A_{ij}} r\left(\vE(\vec X)\right) \mathcal L_{\vec h_L} (\vec X) + r\left(\vE(\vec X)\left(\sigma_{ij}(\vec X)\right)\right) \mathcal L_{\vec h_L}(\sigma_{ij}(\vec X))\; d\vec X =\\
&&\int_{A_{ij}\cup A_{ji}} r\left(\vE(\vec X)\right) \mathcal L_{\vec h_L} (\vec X) \; d\vec X,
\end{eqnarray*}
where the inequality (*) uses the simple inequality for non-negative numbers:
$$ a\geq c, b\geq d \Rightarrow ab + cd  \geq ad + bc.$$
Since $\vec D,\vE$ differ only on $A_{ij}\cup A_{ji}$ and we can repeat this operation for all $i,j$ pairs, this proves property 1.

For property 2, the proof is very similar, if we notice that every constraint for both FWER and FDR  can be written as:
$$Err_{L}(\vec D) = \int_{\Re^K}  s\left(\vec D(\vec X)\right) \mathcal L_{\vec h_L}(\vec X) d\vec X,$$
where $s$ is now arrangement decreasing (because it captures false discovery as opposed to power). The same exact steps replicate the result above, and confirm that property 2 holds for $L=1,\ldots,K-1$. For $L=0$, $s(\vD(\vx))= s(\vE(\vx))$ on $A_{ij}\cup A_{ji}$, since for $\vh=\vec 0$ the function $s(\cdot)$ depends only on the number of ones in the decision vector,  and this number is the same on $A_{ij}\cup A_{ji}$ for $\vD$ or $\vE$: for $Err= FDR, s(\vD) = \frac{\sum_{i=1}^K(1-h_i)D_i}{\vec 1'\vD} \stackrel{\vh=\vec 0}{=}\frac{\sum_{i=1}^KD_i}{\vec 1'\vD}=1$; for $Err=FWER$, $s(\vD) = I(\sum_{i=1}^K(1-h_i)D_i>0)\stackrel{\vh=\vec 0}{=} I(\vec 1'\vD>0)$.

{
\subsection*{Proof of Proposition \ref{prop-existence}}
From Assumption \ref{ass:nonred}, it follows that  
 for any series $\vmu_n$ converging to $\vmu$, the decision vector at $\vmu_n$ will converge to the decision vector at $\vmu$ almost everywhere under the non-redundancy condition. From the dominated convergence theorem  it follows that:
 \begin{eqnarray*}
\lim_{n\rightarrow \infty} &&\int_{Q} \left(\sum_{i=1}^K b_{L,i}(\vec u)D^{\vmu_n}_i(\vec u)\right) d\vec u = \int_{Q} \left(\sum_{i=1}^K b_{L,i}(\vec u)D^\vmu_i(\vec u)\right) d\vec u \\\lim_{n\rightarrow \infty}&&\int_{Q}  \left(\sum_{i=1}^K a_i(\vec u) D^{\vmu_n}_i(\vec u)\right) d\vec u= \int_{Q}  \left(\sum_{i=1}^K a_i(\vec u) D^{\vmu}_i(\vec u)\right) d\vec u, 
\end{eqnarray*}
so the following functions are continuous in $\vmu$:  
 \begin{eqnarray*}&&G_L(\vmu) = \alpha - (\int_{Q} \left(\sum_{i=1}^K b_{L,i}(\vec u)D^\vmu_i(\vec u)\right) d\vec u \mbox{ for }L=0,\ldots,K-1 \\ &&\int_{Q}  \left(\sum_{i=1}^K a_i(\vec u) D^{\vmu}_i(\vec u)\right) d\vec u.
 \end{eqnarray*}
Therefore, the dual objective 
$$L(  \vmu) =   \int_{Q}  \left(\sum_{i=1}^K a_i(\vec u) D^{\vmu}_i(\vec u)\right) d\vec u+\sum_{L=0}^{K-1}\mu_L\times G_L(\vmu)$$ 
is continuous in $\vmu$ and the set $\mathcal C = \{\vmu: G_L(\vmu)\geq 0, \mu_L\geq 0, L=0,\ldots,K-1 \}$ is closed. Moreover, $\mathcal C$ is compact since there exists an upper bound $\vec B$ such that for $\vmu\geq \vec B$, $G_L(\vmu)\geq 0$ for all $L=0,\ldots,K-1$.  The existence of 
$$\vmu^*=\arg\min\{L(  \vmu): \vmu \in \mathcal C \}  $$ follows since the minimum of  a continuous function on a compact set necessarily exits. 
} 

\subsection*{Proof of Proposition \ref{thm_comp}} Among all $\vec D:[0,1]^K \rightarrow \{0,1\}^K$ that satisfy $Err_{\vec h,\theta_A}(\vec D)\leq \alpha$,  $\vec D^*(\theta_0, \theta_A)$  has the highest power (by definition): $\Pi_{\theta_0}(\vec D^*(\theta_0, \theta_A))>\Pi_{\theta_0}(\vec D)$. Therefore, if in addition $
Err_{\vec h,\theta} \left( D^*(\theta_0,\theta_A) \right) \leq \alpha\;,\forall \vec h\in\{0,1\}^K,\theta\in\Theta_A
$, $\vec D^*(\theta_0, \theta_A)$ has the highest power among all potential solutions to \eqref{opt_comp}, i.e., it is the optimal solution of the single objective optimization problem.
\subsection*{Proof of Corollary \ref{coro_comp}} Suppose by contradiction that there exists another $\theta \neq \theta_A$ such that $\Pi_{\theta_0}(\vec D^*(\theta_0, \theta_A))>\Pi_{\theta_0}(\vec D^*(\theta_0, \theta))$. Then   $Err_{\vec h,\theta} (\vec D^*(\theta_0, \theta_A))>\alpha$ for at least one $\vec h \in \{0,1\}^K$, otherwise the definition of $\vec D^*(\theta_0, \theta)$ as the optimal solution to  \eqref{two_theta} is violated. But this contradicts the fact that $
Err_{\vec h,\theta} \left( \vec D^*(\theta_0,\theta_A) \right) \leq \alpha\;,\forall \vec h\in\{0,1\}^K,\theta\in\Theta_A,$ thus proving the corollary.
\subsection*{Proof of Theorem \ref{thm_maximin}}
Define a feasible solution $\vec D$ as one that satisfies  $Err_{\vec h,\theta} (\vec D) \leq \alpha\;,\; \forall \vec h \in \{0,1\}^K,\; \theta \in \Theta_A$.
Since $\vec D^*(\theta_0, \theta_A)$ is the optimal solution to \eqref{opt_comp} by Assumption 1, then for any feasible $D$
\begin{equation}\label{thm_maximin_pf-eq1}
   \min_{\theta \in \Theta_B} \Pi_{\theta}(\vec D)\leq  \Pi_{\theta_0}(\vec D) \leq \pi_{\theta_0}(\vec D^*(\theta_0, \theta_A)).    \nonumber
\end{equation}

Moreover,  since by Assumption 2 $$\min_{\theta \in \Theta_B} \Pi_{\theta}(\vec D^*(\theta_0, \theta_A)) = \Pi_{\theta_0}(\vec D^*(\theta_0, \theta_A)),$$
it follows that $\min_{\theta \in \Theta_B} \Pi_{\theta}(\vec D)$ is upper bounded by $\Pi_{\theta_0}(\vec D^*(\theta_0, \theta_A))$. So:
$$ \max_{\vec D:[0,1]^K \rightarrow \{0,1\}^K, \mbox{feasible}} \min_{\theta \in \Theta_B} \Pi_{\theta}(\vec D) = \Pi_{\theta_0}(\vec D^*(\theta_0, \theta_A)).$$

\section{Distributions that satisfy Assumption \ref{ass:exch}}\label{supp-DistArrInc}

Multivariate densities that have the arrangement increasing (or decreasing) property were studied in \cite{Hollander77}. The list includes the multivariate F distribution, the multivariate Pareto distribution, and the multivariate normal distribution with common variance and common pairwise covariance.

Specifically, if the sample means follow a multivariate normal with a fixed pairwise correlation $\rho$ (as occurs, e.g., when multiple treatment groups are compared to the same control group), the assumption is satisfied. 

Note that Assumption \ref{ass:exch} may be satisfied even when Assumption \ref{ass:monoLR} is violated.
For testing $K$ normal means, Assumption \ref{ass:exch} is satisfied if the sample means have a common variance,  and a common pairwise correlation for all pairs that satisfy $h_i=h_j$. The correlation between pairs of null test-statistics, $\rho_{0,0}$, can be different from that of pairs of non-null test-statistics, $\rho_{1,1}$, or of pairs with exactly one null test statistic, $\rho_{1,0}$. But Assumption \ref{ass:monoLR}  can be violated for normal means if $\rho_{0,0}, \rho_{1,0},$ and $\rho_{1,1}$ differ. Hence the two assumptions are not redundant, even for testing normal means.

\section{The ordering property for K non-exchangeable simple hypotheses testing problems}\label{supp-LRorderinggeneral}

For $\vh$-exchangeable hypotheses, we proved that the LR-ordering property is necessarily maintained and this property implies that the smallest $p$-values are rejected. For independent test statistics, we can relax the assumption of $\vh$-exchangeability and consider more generally the setting of $K$ simple null and alternative hypotheses that need not be the same, so the $p$-value $u_i$ of the $i$th likelihood ratio statistic has a uniform null density,  a nonnull density $g_i$, and the nonnull densities need not be the same (i.e. $g_1, \ldots,  g_K$ can be any nonnull densities for the LR $p$-values). 

We shall consider only decision functions that are {\it symmetric in their $p$-values}, i.e., $\sigma(\vD(\vu)) = \vD(\sigma(\vu))$ for all vectors of $p$-values $\vu$ with mutually independent coordinates and $\sigma \in S_K$. We shall  show that the smallest $p$-values are rejected, so it is enough to search for a solution in the ``lower corner set" $Q$.

 The proof is very similar to the proof of   Theorem \ref{lemma-weakmonotone} so the result is stated here as a corollary. 
\begin{coro}
Assuming independence of the LRs for testing $K$ simple hypotheses testing problems, for any of our considered power and level criteria, the optimal symmetric in-their-$p$-values solution, $\vD^*$, satisfies the $p$-value ordering property, i.e., it rejects the smallest $p$-values:
$$u_i<u_j \Rightarrow D_i^*(\vu)\geq D_j^*(\vu). $$
\end{coro}

{\bf Proof:}
Two key observations follow from the assumption of independence and the consideration of decision rules that are symmetric in their $p$-values.
First, from the assumption of independence it follows that if $u_i<u_j$, and $h_i=1, h_j=0,$ then the joint density of $\vu$, $\mathcal L_{\vec h}(\vu)=\Pi_{l=1}^Kg_l(u_l)^{h_l}$, is greater than the joint density of $\sigma_{ij}(\vu)$, where $\sigma_{ij}$ is the permutation that switches the $p$-values for hypotheses $i,j$ only. This follows since the alternative densities are decreasing, as formally stated for completeness in the  Lemma \ref{lemma-LRpvaluedecreasingdensity} below (though this result may be well known), so $g_i(u_i)\geq g_i(u_j)$ for $u_i<u_j$. Therefore, the Arrangement increasing property in Assumption 2.2 is satisfied. 

Second,  defining
$$ A_{ij} =\left\{\vu :  u_i< u_j,\;  D_i(\vu) =0, D_j(\vu)=1 \right\},$$
then by the symmetry requirement of $\vec D$, there is a symmetric set
$$ A_{ji} =\left\{ \vu :  u_j< u_i,\;  D_j(\vu) =0, D_i(\vu)=1 \right\},$$
where
$\vu \in A_{ij} \Leftrightarrow \sigma_{ij}(\vu) \in A_{ji}.$

As in the proof of Theorem \ref{lemma-weakmonotone} we  examine the solution $\vE$ which is equal to $\vec D$ everywhere, except on the sets $A_{ij}, A_{ji}$, where it switches the value of coordinates $i,j$:
$$
E_k(\vu) = \left\{ \begin{array}{ll} D_k(\vu) & \mbox{if } \vu \notin A_{ij}\cup A_{ji} \mbox{ or } k\notin \{i,j\}\\
D_j(\vu) & \mbox{if } \vu \in A_{ij}\cup A_{ji} \mbox{ and } k=i \\
D_i(\vu) & \mbox{if } \vu \in A_{ij}\cup A_{ji} \mbox{ and } k=j
\end{array} \right. 
$$

For any function $a(\cdot),$ we have from symmetry that:   \begin{eqnarray}
&\int_{A_{ij}\cup A_{ji}}& a\left(\vec D(\vu)\right) \mathcal L_{\vec h}(\vu)\;  d\vu  = \int_{A_{ij}}  a\left(\vec D(\vu)\right) \mathcal L_{\vec h} (\vu) + a\left(\vec D\left(\sigma_{ij}(\vu)\right)\right) \mathcal L_{\vec h}\left(\sigma_{ij}(\vu)\right)\; d\vu \nonumber \\ \label{eq1-nonexch}
\end{eqnarray}

 The proof  proceeds with the exact same reasoning as  in the proof of Theorem \ref{lemma-weakmonotone}. Specifics follow for showing that for all constraints we consider $Err_{\vh}(\vE)\leq Err_{\vh}(\vD)$.
We can write the error as $Err_{\vh}=\int_{[0,1]^K} a(\vD)\mathcal L_{\vec h}(\vu)\;  d\vu$, where $a(\vD)$ is arrangement decreasing in $\vD$. For FDR, $a(\vD) = \frac{\sum_{l=1}^K (1-h_l)D_l}{\sum_{l=1}^K D_l}$, so if $h_i=h_j$ or $D_i=D_j$ then $a(\vD(\vu)) = a(\vD(\sigma_{ij}(\vu))$, but if $h_i=1, h_j=0$ and $D_i=0, D_j=1$ then $$a(\vD(\sigma_{ij}(\vu))-a(\vD(\vu)) =  \frac{-1}{\sum_{l=1}^K D_l}\leq 0.$$ For FWER, $a(\vD) = I(\sum_{l=1}^K (1-h_l)D_l>0)$, so if  $h_i=1, h_j=0$ and $D_i=0, D_j=1$, then $$a(\vD(\sigma_{ij}(\vu))-a(\vD(\vu)) =  I\left(\sum_{l=1, l\neq i, l\neq j}^K (1-h_l)D_l>0\right)-1\leq 0. $$
Note that for both of these functions, by construction, when $\vu \in A_{ij} \cup  A_{ji}$ then $a(\vD(\vu)) = a(\vE(\sigma_{ij}(\vu)))$ and  
$a(\vE(\vu)) = a(\vD(\sigma_{ij}(\vu))).$

Using  the simple inequality for non-negative numbers,
$ a\geq c, b\geq d \Rightarrow ab + cd  \geq ad + bc$, it follows that 
 for the quantity in \eqref{eq1-nonexch} we have:  
\begin{eqnarray*}
&&\int_{A_{ij}\cup A_{ji}} a\left(\vec D(\vu)\right) \mathcal L_{\vec h}(\vu)\;  d\vu \geq\\
&& \int_{A_{ij}} a\left(\vec D\left(\sigma_{ij}(\vec u)\right)\right) \mathcal L_{\vec h} (\vu) + a\left(\vec D(\vec u)\right) \mathcal L_{\vec h}\left(\sigma_{ij}(\vec u)\right)\; d\vec u  =\\ &&\int_{A_{ij}} a\left(\vE(\vec u)\right) \mathcal L_{\vec h} (\vec u) + a\left(\vE\left(\sigma_{ij}(\vec u)\right)\right) \mathcal L_{\vec h}(\sigma_{ij}(\vec u))\; d\vec u =\\
&&\int_{A_{ij}\cup A_{ji}} a\left(\vE(\vec u)\right) \mathcal L_{\vec h} (\vec u) \; d\vec u.
\end{eqnarray*}
Since $\vec D,\vE$ differ only on $A_{ij}\cup A_{ji}$ and we can repeat this operation for all $i,j$ pairs, this proves that for all constraints we consider, $Err_{\vh}(\vE)\leq Err_{\vh}(\vD)$.

\begin{lem}\label{lemma-LRpvaluedecreasingdensity}
For testing a simple hypothesis against a simple alternative, the density of the $p$-value of the LR test statistic is decreasing.  
\end{lem}
{\bf Proof:}
Let $T$ be the LR test statistic, with density and cummulative density $f_0, F_0$ under the null hypothesis $H_0$ and $f_1, F_1$ under the alternative hypothesis $H_1$. The $p$-value is $u=1-F_0(t)$, for $t$ the realized value of $T$. In order to derive the non-null density of the $p$-value, note that $$\mP_{H_1}(U\leq u) = \mP_{H_1}(1-F_0(T)\leq u) = \mP_{H_1}\left(T\geq F_0^{(-1)}(1-u)\right) = 1-F_1\left(F_0^{(-1)}(1-u)\right). $$
The nonnull density of the $p$-value is therefore
$$
g(u) = \frac{d\left \lbrace 1-F_1\left(F_0^{(-1)}(1-u)\right)\right \rbrace }{du} = \frac{f_1\left (F_0^{(-1)}(1-u)\right)}{f_0\left(F_0^{(-1)}(1-u)\right)} = \frac{f_1(t)}{f_0(t)}. 
$$
Since $u_1\leq u_2 \iff T_1>T_2$, it remains to show that $\frac{f_1(t)}{f_0(t)}$ is increasing in $t$ (thus implying that $g(u_1) >g(u_2)$). This follows since $f_1(t) = t\times f_0(t). $ To see this, note that for sample $\vx$ with  likelihood $\mathcal L_0(\vx)$ under the null and  $\mathcal L_1(\vx)$ under the alternative,
\begin{eqnarray}
&& F_1(t)  = \int_{\lbrace \vx: T(\vx)\leq t \rbrace }\mathcal L_1(\vx)d\vx = \int_{\lbrace \vx: T(\vx)\leq t \rbrace }T(\vx) \mathcal L_0(\vx)d\vx\nonumber  \\ 
&& = \mE_0 \left \lbrace T\times I(T\leq t)\right \rbrace = \int_0^t \lambda f_0(\lambda) d\lambda \nonumber
\end{eqnarray}
Differentiating, we get that $$f_1(t)  = \frac{d\int_0^t \lambda f_0(\lambda) d\lambda}{dt}= t\times f_0(t).  $$


\section{Derivation of the expression for $FDR_L(D)$}\label{supp-FDRexpression}
If $L=0$, then $FDR_0(\vec D) = FWER_0(\vec D) = K!\int_Q D_1(\vec u)d\vec u$.
Let $FDP_k$ denote the false discovery proportion if the $k$ smallest $p$-values are rejected, i.e., if $D_k(\vec u)-D_{k+1}(\vec u)=1$.
For $L>0$, the expression \eqref{FDR-const} in the main manuscript follows since:
\begin{eqnarray}
&& FDR_L(\vec D) = \mE\left\lbrace \sum_{k=1}^{K-1}[D_k(\vec u)-D_{k+1}(\vec u)]FDP_k + D_K(\vec u) FDP_K \right \rbrace \nonumber\\
&& = \mE\left\lbrace \sum_{k=1}^K D_k(\vec u)FDP_k -\sum_{k=1}^{K-1} D_{k+1}(\vec u)FDP_k \right \rbrace \nonumber\\
&& = \mE\left\lbrace \sum_{k=1}^K D_k(\vec u)FDP_k -\sum_{k=2}^{K} D_{k}(\vec u)FDP_{k-1} \right \rbrace \nonumber\\
&& = \mE\left\lbrace D_1(\vec u)FDP_1+\sum_{k=2}^K D_k(\vec u)[FDP_k -FDP_{k-1}] \right \rbrace \nonumber\\
&& =   \int_{[0,1]^K}   f_{\vec h_L}(\vec u) \left\lbrace D_1(\vec u)FDP_1+\sum_{k=2}^K D_k(\vec u)[FDP_k -FDP_{k-1}] \right \rbrace
     d\vec u\nonumber\\
&& =   \int_{[0,1]^K}  \frac{\sum_{i \in \binom{K}{L}} f_i(\vec u)}{\binom{K}{L}} \left\lbrace D_1(\vec u)FDP_1+\sum_{k=2}^K D_k(\vec u)[FDP_k -FDP_{k-1}] \right \rbrace
     d\vec u\nonumber\\
&& =   \int_Q   L!(K-L)!\sum_{i \in \binom{K}{L}} f_i(\vec u) \left\lbrace D_1(\vec u)FDP_1+\sum_{k=2}^K D_k(\vec u)[FDP_k -FDP_{k-1}] \right \rbrace
     d\vec u
, \nonumber \\\label{supp-FDRexpression-eq1}
\end{eqnarray}
where the next to last equality follows from the exchangeability assumption, and the last equality follows from the additional symmetry assumption.

For $i \in \binom{K}{L}$, the difference $FDP_k-FDP_{k-1}$ for $k>1$ reduces to
$$ r_{ki} = \frac{\left|\{1,\ldots\,k\} \cap i^c\right|}{k} - \frac{\left|\{1,\ldots\,k-1\} \cap i^c\right|}{k-1},$$
    where $i^c$ denotes the actual set of true nulls in the configuration indexed by $i$. Setting  $0/0 = 0$, $FDP_1 = r_{1i}$. Therefore expression \eqref{supp-FDRexpression-eq1}  is equal to:
     \begin{eqnarray}
&&         \int_Q   L!(K-L)!\sum_{i \in \binom{K}{L}} f_i(\vec u) \left\lbrace D_1(\vec u)r_{1i}+\sum_{k=2}^K D_k(\vec u)r_{ki} \right \rbrace
     d\vec u  \nonumber\\
     && = L!(K-L)! \int_Q \sum_k D_k(\vec u)
    \sum_{i \in \binom{K}{L}} f_i(\vec u) r_{ki}  d\vec u. \nonumber
     \end{eqnarray}

\section{Maximizing $\Pi_3$ for $K=3$ independent tests under FDR control} \label{supp:alg}
We now choose a specific instance of the general problem in \S~\ref{Sec:main-res} of the main manuscript, to demonstrate a detailed derivation of the formulas and the resulting algorithm.
We use the power function $\Pi_{L=3}$, that is, maximizing average power in case all nulls are in fact false. Recall that $g$ denotes the density of each coordinate of $\vec u$ under the alternative.

\newcommand{\tnt}{\ensuremath{\mbox{int}}}

Putting together the objective and constraints $FDR_L(\vec D)$ from Eq. \eqref{FDR-const} in the main manuscript for the relaxed infinite linear program on $Q$ gives:
\begin{eqnarray*}
\max_{\vec D: Q\rightarrow[0,1]^3} && \int_Q \left(D_1(\vec u)+D_2(\vec u)+D_3(\vec u)\right)  g(u_1)g(u_2)g(u_3) d\vec u  \nonumber \\
\label{L0} \mbox{s.t. } &&   FDR_0(\vec D)=6 \int_Q  D_1(\vec u) d\vec u \leq \alpha,   \\
\label{L1} && FDR_1(\vec D)=2 \int_Q  \left[ D_1(\vec u) \left(g(u_2)+g(u_3)\right) +  D_2(\vec u) \frac {g(u_1)-g(u_2)}{2}  +\right.  \\ &&\;\;\;\;\;\; \left. D_3(\vec u) \frac {g(u_1)+g(u_2)-2g(u_3)}{6}\right]  d\vec u \leq \alpha,  \nonumber \\
\label{L2}  && FDR_2(\vec D)=2 \int_Q  \left[ D_1(\vec u) g(u_2)g(u_3) + D_2(\vec u) \frac{g(u_1)g(u_3)-g(u_2)g(u_3)}{2} +\right. \\ &&\;\;\;\;\; \left.D_3(\vec u) \frac{2g(u_1)g(u_2)-g(u_1)g(u_3)-g(u_2)g(u_3)}{6} \right] d\vec u \leq \alpha,  \nonumber \\
&&0 \leq D_3(\vec u) \leq D_2(\vec u) \leq D_1(\vec u) \leq 1,\; \forall \vec u \in Q, \nonumber
\end{eqnarray*}

Applying our results we denote:
\begin{eqnarray*}
R_1(\vec u) = &&g(u_1)g(u_2)g(u_3) - 6\mu_0 - 2\mu_1 \left( g(u_2) + g(u_3) \right)-   2\mu_2 g(u_2)g(u_3)\\
R_2(\vec u) = && g(u_1)g(u_2)g(u_3)  - 2\mu_1 \frac{ g(u_1) - g(u_2)}{2}  - \\  &&2\mu_2 \frac{g(u_1)g(u_3)-g(u_2)g(u_3)}{2} \\
R_3(\vec u) = && g(u_1)g(u_2)g(u_3)  - 2\mu_1 \frac{g(u_1)+g(u_2)-2g(u_3)}{6} - \\  &&2\mu_2 \frac{2g(u_1)g(u_2)-g(u_1)g(u_3)-g(u_2)g(u_3)}{6},
\end{eqnarray*}
and use these to define the corresponding $D$ functions:
\begin{eqnarray*}
D^{\vmu}_1(\vec u) = && \mathbb I\left\{ R_1(\vec u)> 0 \cup R_1(\vec u)+R_2(\vec u)>0 \cup R_1(\vec u)+R_2(\vec u)+R_3(\vec u) > 0 \right\}\\
D^{\vmu}_2(\vec u) = && \mathbb I\left\{ D^{\vmu}_1(\vec u) \cap (R_2(\vec u)> 0 \cup R_2(\vec u)+R_3(\vec u)>0 \right\}\\
D^{\vmu}_3(\vec u) = && \mathbb I\left\{ D^{\vmu}_2(\vec u) \cap R_3(\vec u)> 0 \right\},
\end{eqnarray*}
and use these to search for $\vmu^*$ complying with Eq. (\ref{eq:mucond}) in the main manuscript.

\section{Rejection regions for maximizing $\Pi_3$ for $K=3$ independent tests with FWER control and FDR control}\label{supp:subsec-3FWER}

\begin{figure}[htbp]
  \begin{tabular}{cccc}
  $u_1=1.59e-05$ & $u_1=1.66e-02$ & $u_1=4.38e-02$ & $u_1=5.38e-02$\\
  \includegraphics[width=3cm,height=3cm,page=1]{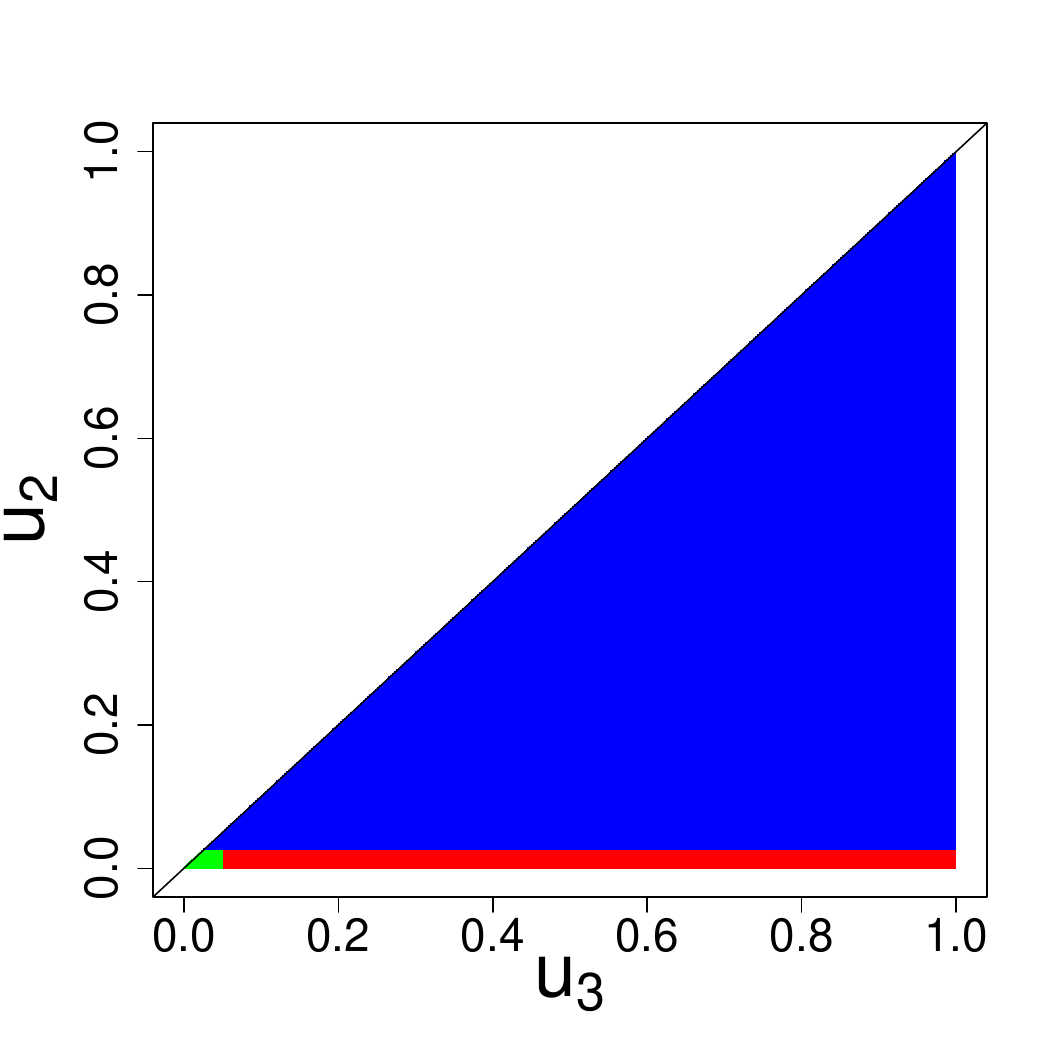} &
  \includegraphics[width=3cm,height=3cm, page=2]{FWER.pdf} &  &
\\
\includegraphics[width=3cm,height=3cm, page=3]{FWER.pdf}  &
\includegraphics[width=3cm,height=3cm, page=4]{FWER.pdf}  &
\includegraphics[width=3cm,height=3cm, page=5]{FWER.pdf}  &
\includegraphics[width=3cm,height=3cm, page=6]{FWER.pdf}
\\
\includegraphics[width=3cm,height=3cm, page=7]{FWER.pdf} &
\includegraphics[width=3cm,height=3cm, page=8]{FWER.pdf} &
\includegraphics[width=3cm,height=3cm, page=9]{FWER.pdf} &
\includegraphics[width=3cm,height=3cm, page=10]{FWER.pdf}
\\
 \includegraphics[width=3cm,height=3cm, page=11]{FWER.pdf}& \includegraphics[width=3cm,height=3cm, page=12]{FWER.pdf}&
 \includegraphics[width=3cm,height=3cm, page=13]{FWER.pdf}&
 \includegraphics[width=3cm,height=3cm, page=14]{FWER.pdf}
 \\
\includegraphics[width=3cm,height=3cm, page=15]{FWER.pdf} &
\includegraphics[width=3cm,height=3cm, page=16]{FWER.pdf} &
\includegraphics[width=3cm,height=3cm, page=17]{FWER.pdf} &
\includegraphics[width=3cm,height=3cm, page=18]{FWER.pdf}
\\
 \includegraphics[width=3cm,height=3cm, page=19]{FWER.pdf}& \includegraphics[width=3cm,height=3cm, page=20]{FWER.pdf}&
 \includegraphics[width=3cm,height=3cm, page=21]{FWER.pdf}&
 \includegraphics[width=3cm,height=3cm, page=22]{FWER.pdf}
 \end{tabular}
 \caption{\small For fixed values of the minimum $p$-value, the 2-dimensional slices of the following 3-dimensional rejection regions for strong FWER control at level 0.05: Bonferroni-Holm (row 1); OMT procedure  for $\Pi_{\theta,3}$ at $\theta=-0.5$ (row 2),  at $\theta = -1.33$ (row 3), and at $\theta = -2$ (row 4); OMT procedure  for $\Pi_{\theta,any}$ at $\theta=-1.33$ (row 5), and for any $\theta>-0.75$ or $\theta<-1.6$ (row 6). In green: reject all three hypotheses; in red: reject exactly two hypotheses; in blue: reject only one hypothesis. Since Bonferroni-Holm makes no rejections if all $p$-values are greater than 0.05/3, the top two right panels are empty. For each panel, the rejection region is in the top right quadrant of the partition of the plane by the point $(u_1, u_1)$.}
 \label{fig3FWER}
\end{figure}

The rejection regions of the different procedures presented in \S~\ref{Sec:main-res}
 of the main manuscript for FWER control with 3 independent normal means are displayed in Figures \ref{fig3FWER}. For a 2-dimensional display, we selected slices of the 3-dimensional rejection region that are fixed by the minimum $p$-value.
We show for different procedures four slices from policies with FWER control: the slices with a very small minimum $p$-value, the largest minimum $p$-value for which Bonferroni-Holm still makes rejections (i.e., 0.05/3), a minimal $p$-value slightly below the nominal level, and a  minimal $p$-value above the nominal level.

We shall now consider the optimal FDR controlling procedure for the $K=3$ independent normal means. 
The following  procedure, which is  uniformly more powerful than the BH procedure, was suggested in \cite{Solari17}. If  there exists an $i\in \{1,\ldots,K \}$ such that $u_i\leq i\alpha/K$, then reject all hypotheses up to $\max\{i: u_i\leq i\alpha/(K-1)\}$. They called this procedure minimally adaptive Benjamini-Hochberg (MABH).

\begin{figure}[htbp]
  \begin{tabular}{cccc}
  $u_1=1.59e-05$ & $u_1=1.73e-02$ & $u_1=4.38e-02$ & $u_1=1.06e-1$\\
  \includegraphics[width=3cm,height=3cm, page=1]{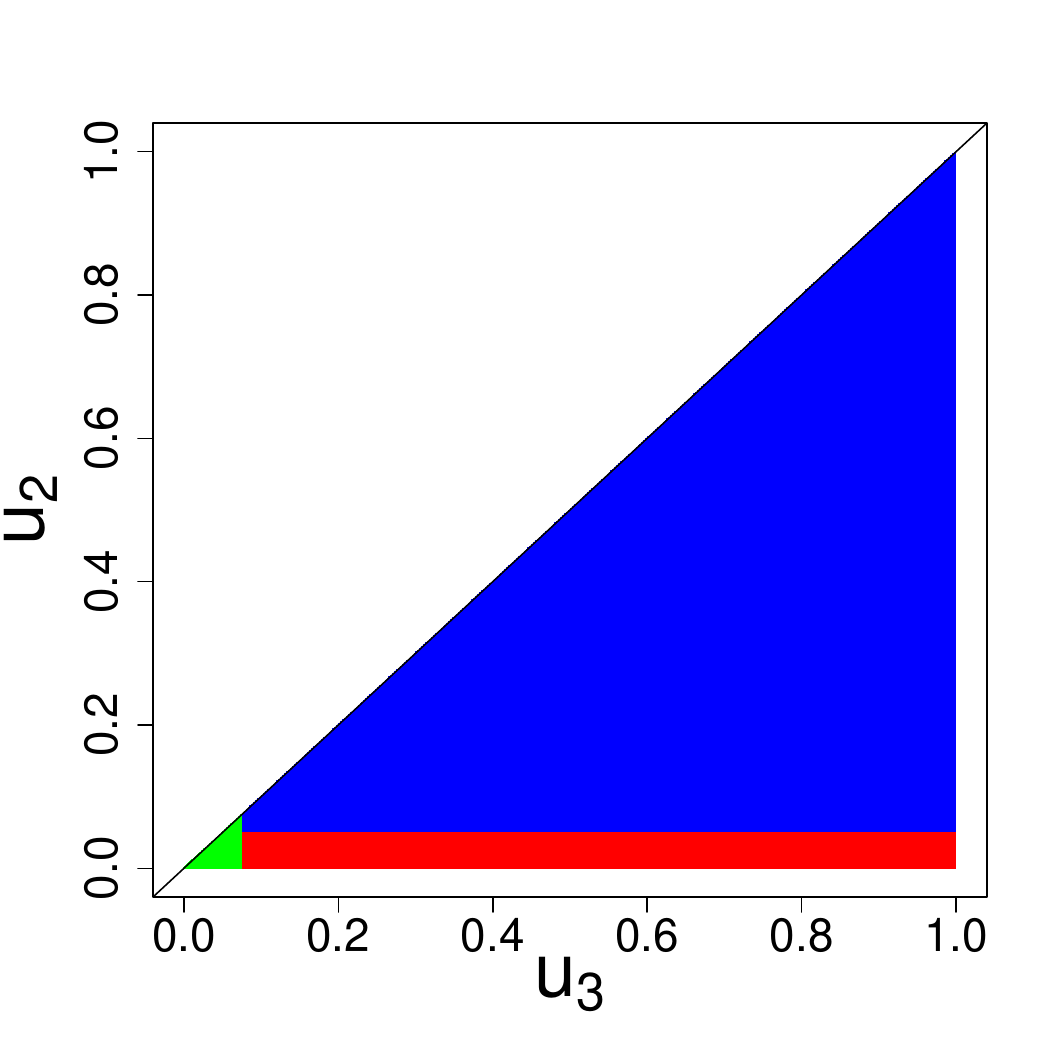} &
  \includegraphics[width=3cm,height=3cm, page=2]{FDR.pdf} &
  \includegraphics[width=3cm,height=3cm, page=3]{FDR.pdf} &
\\
\includegraphics[width=3cm,height=3cm, page=4]{FDR.pdf} &
\includegraphics[width=3cm,height=3cm, page=5]{FDR.pdf} &
\includegraphics[width=3cm,height=3cm, page=6]{FDR.pdf} &
\includegraphics[width=3cm,height=3cm, page=7]{FDR.pdf}
\\
\includegraphics[width=3cm,height=3cm, page=8]{FDR.pdf}  &
\includegraphics[width=3cm,height=3cm, page=9]{FDR.pdf}  &
\includegraphics[width=3cm,height=3cm, page=10]{FDR.pdf}  &
\includegraphics[width=3cm,height=3cm, page=11]{FDR.pdf}
\\
 \includegraphics[width=3cm,height=3cm, page=12]{FDR.pdf}& \includegraphics[width=3cm,height=3cm, page=13]{FDR.pdf}&
 \includegraphics[width=3cm,height=3cm, page=14]{FDR.pdf}&
 \includegraphics[width=3cm,height=3cm, page=15]{FDR.pdf}
 \end{tabular}
 \caption{\small For fixed values of the minimum $p$-value, the 2-dimensional slices of the following 3-dimensional rejection regions for strong FDR control at level 0.05: MABH (row 1); OMT rejection regions for $\Pi_{\theta,3}$ at $\theta=-0.35$ (row 2),  at $\theta = -0.5$ (row 3), and at $\theta = -2$ (row 4). In green: reject all three hypotheses; in red: reject exactly two hypotheses; in blue: reject only one hypothesis. Since MABH makes no rejections if all $p$-values are greater than 0.05, the top right panel is empty. For each panel, the rejection region is in the top right quadrant of the partition of the plane by the point $(u_1,u_1)$. 
  }
 \label{fig3FDR}
\end{figure}

Both BH and MABH have monotone decision rules (Def. \ref{def-monotone}), and therefore may have reduced power in comparison to  procedures that violate the monotonicity property, in particular OMT procedures. For $K=3$ hypotheses, Table \ref{tab-power-FDR} shows the power comparison for various values of $\theta$. We see that OMT policies optimized for $\Pi_{\theta,3}$ are significantly more powerful than BH and MABH: they offer more than three-fold power for the lower-power settings $\theta\in \{-0.35,-.5\}$ and about 25\% more power in the higher power setting.  From a comparison of rejection regions in Figure \ref{fig3FDR}, we see that 
 the rejection region near the diagonal is  larger than with MABH, and rejections of hypotheses with $p$-values greater than $\alpha$ are possible. As expected, the OMT rejection regions are not monotone. The non-monotone behaviour separating rejections from no rejections is reasonable. However, a less reasonable  non-monotone behaviour of the optimal rejection regions is the positive slope separating the red and green regions  for $\theta=-2$.
Such a rejection region is counter-intuitive, since it contradicts the reasonable principle that if $(u_1,u_2,u_3)\leq (u_1',u_2',u_3')$ elementwise and $(u_1',u_2',u_3')$ result in rejection of all null hypotheses, then $(u_1,u_2,u_3)$ will also result in rejection of all null hypotheses.

\begin{table}[ht]
\centering
\caption{Average power (i.e., probability of rejecting each hypothesis) for various policies with strong FDR control guarantee at the 0.05 level.   }\label{tab-power-FDR}
\begin{tabular}{|cccc|}
  \hline
$\theta$ &   BH  & MABH &  OMT policy for $\Pi_{\theta,3}$ \\ \hline
-0.35 & 0.042 & 0.045 & 0.150 \\
-0.5 &  0.059 & 0.064 & 0.196\\
-2 &  0.574 & 0.633 & 0.799 \\ \hline
\end{tabular}
\end{table}

For weak signal ($\theta=-0.35$),  the only tight constraint is the global null constraint. Since we are maximizing the average power, the OMT procedure is to reject all three null hypotheses if  $\sum_{i=1}^3\Phi^{-1}(u_i)/\sqrt 3\leq z_{\alpha}$.

We can find the range of values of $\theta$ for which the rejection policy that rejects all three hypotheses if the global null is rejected is valid. It satisfies  $FDR_L = \frac{3-L}{3}\Phi( z_\alpha- L\theta/\sqrt{3})$ for $L=1,2$. For $\alpha=0.05$, $FDR_1\leq \alpha$ if $\theta \geq -0.356$, and $FDR_2\leq \alpha$ if $\theta\geq -0.527$. Therefore, this policy achieves strong FDR control whenever $\theta \geq -0.356$. This policy is clearly optimal for $\theta = -0.35$ since it is the optimal policy if the only constraint is the global null constraint. The following proposition provides the general result.

\begin{prop}\label{sup:prop-FDR}
Let $\theta^* = \min\{\theta: \frac{K-L}{K}\Phi(z_\alpha- L\theta/\sqrt{K})\leq \alpha \ \forall \ L=1,\ldots,K-1\}$. Then for $\theta\geq \theta^*$, the OMT policy for $\Pi_{\theta,K}$ with strong FDR control at level $\alpha$ is to reject all $K$ hypotheses if
$$\sum_{i=1}^K \Phi^{-1}(u_i)/\sqrt K<z_{\alpha}.$$
\end{prop}

In our example, this result applies at $\theta=-0.35$. For $\theta=-0.5$, the only tight constraint is $FDR_1= 0.05$, and the optimal rejection region  includes either three or one rejections. For $\theta=-2$, the tight constraints are $FDR_1=FDR_2=0.05$ (the global null constraint is loose), and either one, two or three rejections can occur.

\section{The effect of miss-specifications on the OMT policy for  simple alternatives}\label{supp:mispec}

In the three normal means setting detailed in the previous section
, we demonstrated in examples that the power gain of the OMT procedure can be large. We examine here the effect on power of miss-specifying the $\theta$ value for which the rejection policy was optimized, as well as the effect on power and validity of miss-specifying the dependence structure (i.e., assuming independence in construction of the OMT policy, while in fact the test statistics are dependent).   Specifically, we show the results  of miss-specification on power and validity, for the following OMT policy: optimizing $\Pi_{-1.33,3}$ with strong FWER control, assuming the three test statistics are independent. Four slices of the OMT region are depicted in row 3 of Figure \ref{fig3FWER}.

\begin{figure}[htbp]
 \includegraphics[width=8cm,height=8cm,page=1]{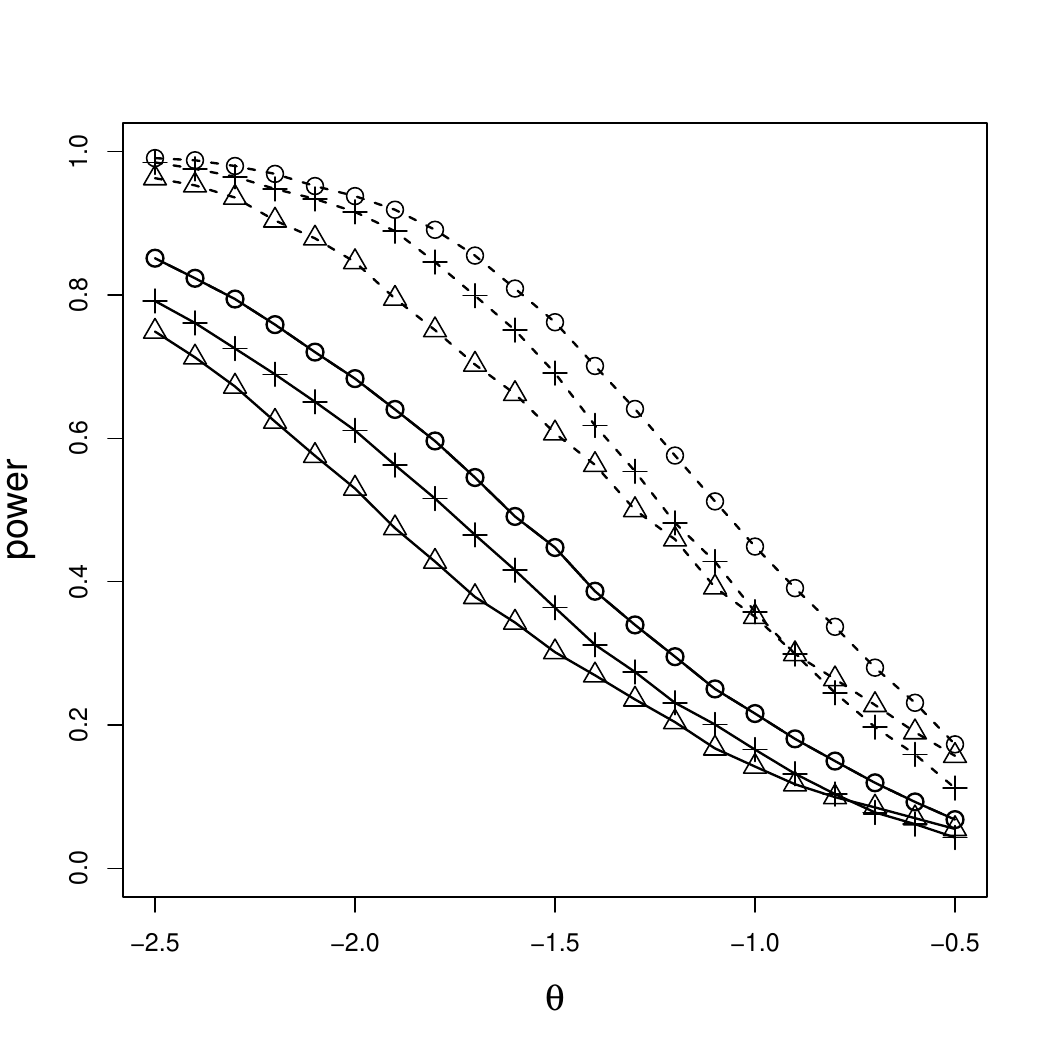}  \caption{\small Average power (solid) and power to reject at least one (dashed) versus signal strength $\theta$ for all three testing procedures. The three procedures are: Bonferroni-Holm/Sidak (triangles); closed-Stouffer (plus signs);  the OMT policy for $\Pi_{-1.33,3}$ (circles).  }
 \label{supfig3mispecPower}
\end{figure}

By optimizing at the wrong $\theta$ value, the power advantage over closed-Stouffer and Bonferroni-Holm/Sidak is maintained when indeed all three hypotheses are nonnull with signal strength $\vec \theta = (\theta, \theta,\theta)$ for $\theta<0$ (Figure \ref{supfig3mispecPower}). Even though optimized at $\theta=-1.33$ for average power, the probability of at least one rejection, as well as the average number of rejections, is largest for this OMT policy for all $\theta$ values. Of course, this comes at a price of the theoretical guarantees being weaker than the guarantees of the other two procedures. The underlying assumptions for the OMT policy are independence of test statistics (as with closed-Stouffer), as well as that the test statistics come from a Gaussian distribution, and the FWER control guarantee is only at $(0,0,0)$, $(0,-1.33,0)$, $(0, -1.33, -1.33)$. Since the exact  knowledge of the parameter value under the alternative is rarely known, we suggest in practice to  use the maximin procedure detailed in \S~\ref{Sec:maximin} of the main manuscript instead. We demonstrate there  that the power advantage is still maintained, now with the same theoretical guarantee as with closed-Stouffer and Bonferroni-Holm/Sidak, as long as the alternatives are indeed Gaussian.

\begin{figure}[htbp]
  \begin{tabular}{ccc}
 \hspace{-1cm}
  \includegraphics[width=5cm,height=5cm, page=1]{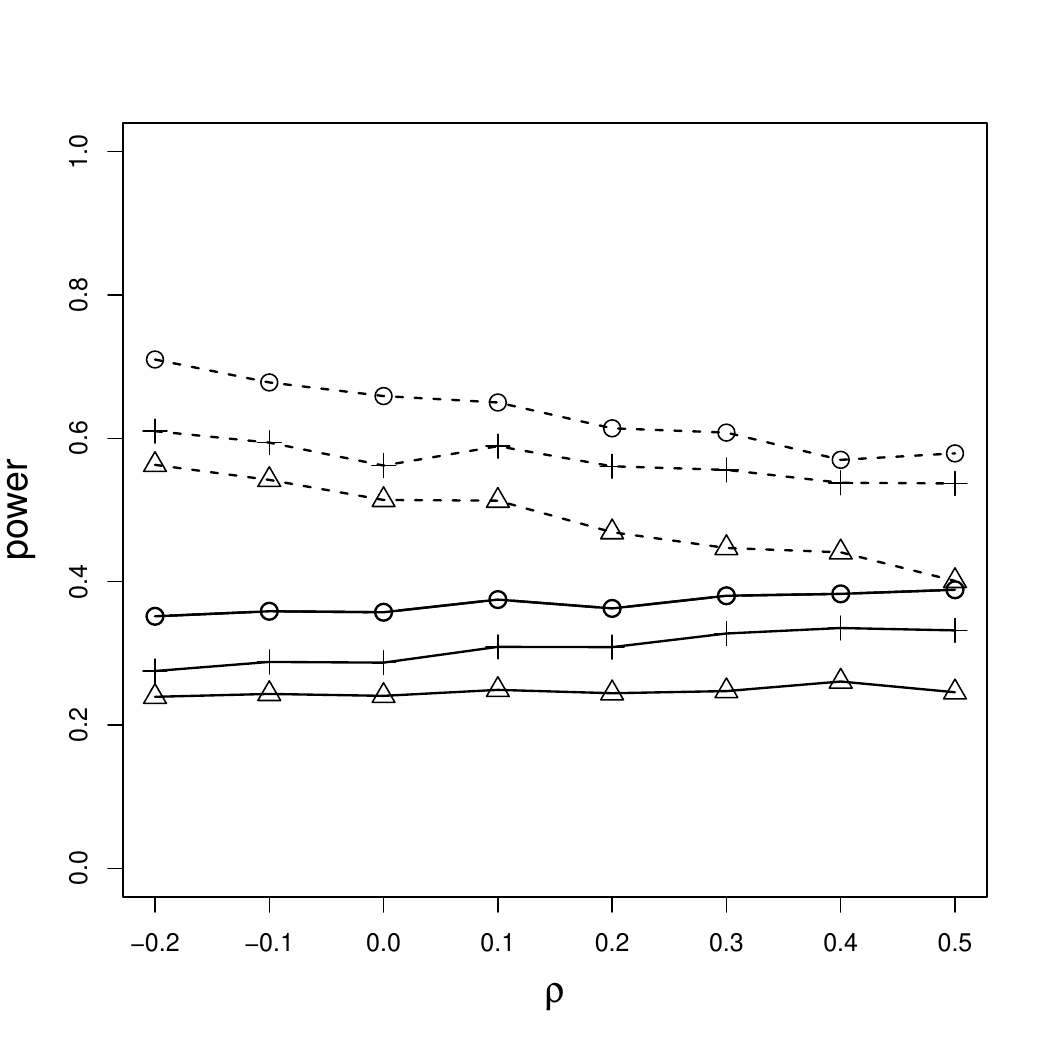} &
  \includegraphics[width=5cm,height=5cm, page=1]{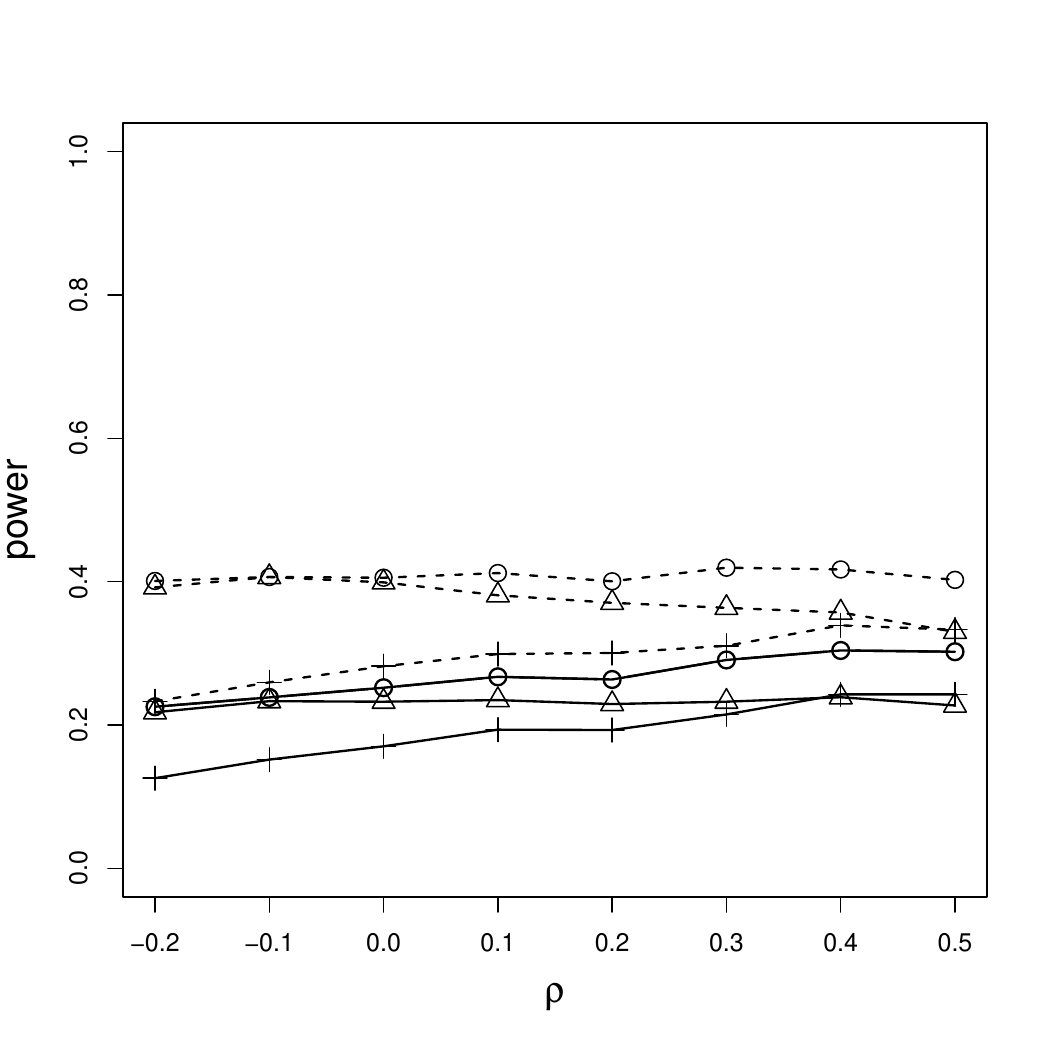} &
  \includegraphics[width=5cm,height=5cm, page=1]{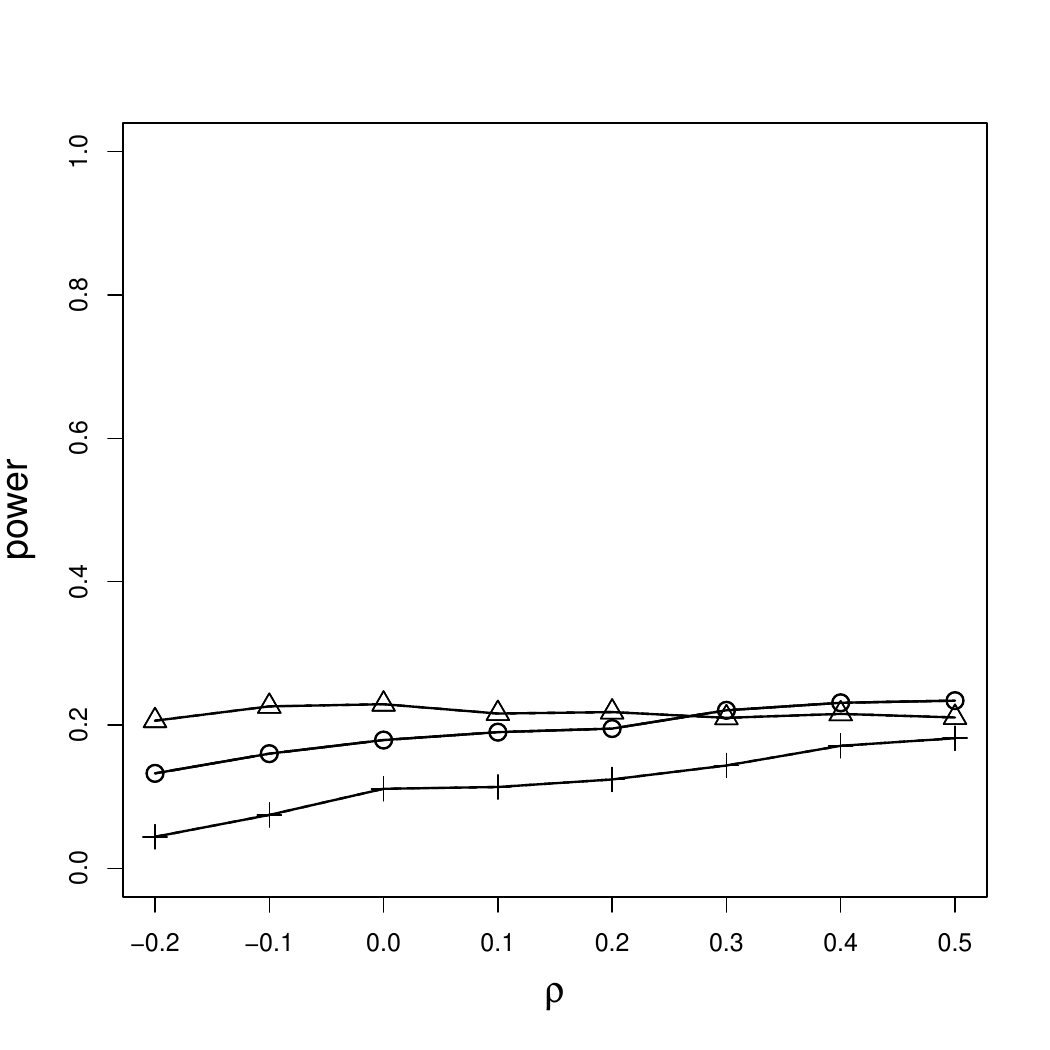}
\\
 \hspace{-1cm}
\includegraphics[width=5cm,height=5cm, page=1]{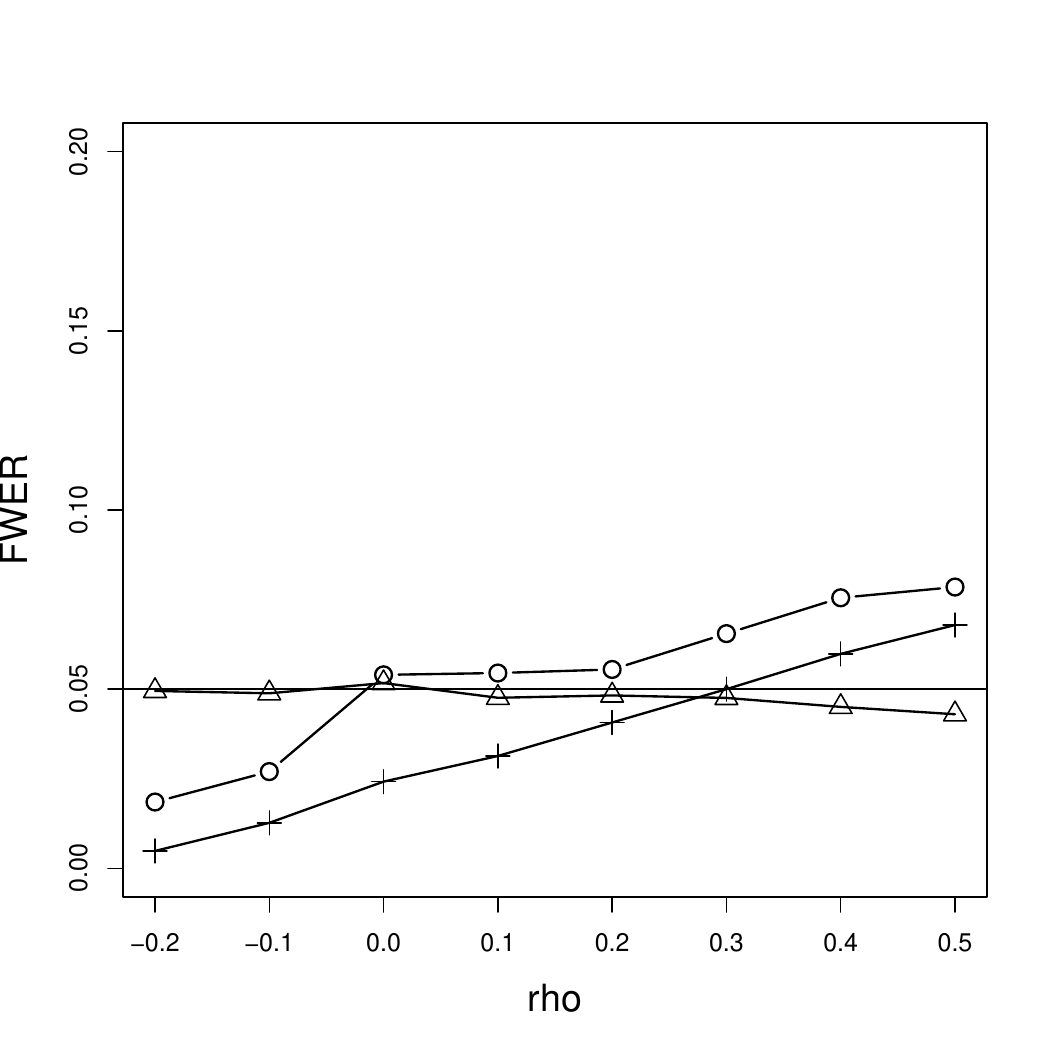}
&
\includegraphics[width=5cm,height=5cm, page=2]{smfigdep-twononnulls.pdf} &
\includegraphics[width=5cm,height=5cm, page=2]{smfigdep-onenonnulls.pdf}

 \end{tabular}
 \caption{\small Power (top row) and FWER (bottom row) versus correlation strength. The three procedures are: Bonferroni-Holm/Sidak (triangles); closed-Stouffer (plus signs);  the OMT policy for $\Pi_{-1.33,3}$ (circles). The panels are: the average power (solid) and power to reject at least one (dashed) when $\vec \theta = (-1.33,-1.33,-1.33)$ (top left panel), $\vec \theta = (0,-1.33,-1.33)$ (top middle panel), and $\vec \theta = (0,0,-1.33)$ (top right panel) for all three testing procedures; the FWER when $\vec \theta = (0,0,0)$ (bottom left panel), $\vec \theta = (0,-1.33,-1.33)$ (bottom middle panel), and $\vec \theta = (0,0,-1.33)$ (bottom right panel) for all three testing procedures.  }
 \label{supfig4mispecPower}
\end{figure}

By assuming independence, even though the test statistics are dependent  with equal correlation $\rho$,  the power advantage over closed-Stouffer and Bonferroni-Holm/Sidak is maintained when  two, or three hypotheses are nonnull with signal strength $\theta=-1.33$, but Bonferroni-Holm/Sidak is superior in power when only a single hypothesis is nonnull (top row of Figure \ref{supfig4mispecPower}). The FWER is inflated when the correlation is positive and above 0.1, and the inflation increases with correlation strength (bottom row of Figure \ref{supfig4mispecPower}). This shows that ignoring the dependence across test statistics can easily result in a non-valid procedure.  Therefore, if the dependence is known (e.g., when multiple treated groups are compared to the same control), it is important to find the OMT policy with the known dependence. For unknown dependence, it may be possible to  estimate the correlation from the data. A careful investigation of the performance of the OMT policy after dependence estimation from the data is outside the scope of this work.

\section{Efficiently computing equations \eqref{FWER-const}-\eqref{FDR-const}}\label{supp:dyn}
\subsection{Dynamic Programming for exponential sums}
\label{dp_basic}
For equations \eqref{FWER-const} and \eqref{FDR-const} we are required to compute sums of an exponential number of functions of the general form $\sum_{i\in {K\choose L}} f_i(u)$, with additional factors or constraints embedded into the summation.

In this subsection we will concentrate on the basic summation without additional factors or constraints, and show that it can be computed efficiently using dynamic programming under the i.i.d. assumption. Under this assumption, $f_i(u)$ can be written as a product of the marginal functions:
\begin{equation}
f_i(u)=\Pi_{j=1}^K f_{i_j}(u_j),
\end{equation}
Where $f_0(u_j)$ and $f_1(u_j)$ the marginal distributions for true null and false null respectively.

we will now break the original problems into smaller subproblems  that will allow us to employ the dynamic programming technique:
\begin{equation}
a_{x,y,z}=\sum_{i\in {{y-x+1}\choose z}} \Pi_{j=1}^{y-x+1} f_{i_j}(u_{x+j-1}).
\end{equation}
The sum computed in $a_{x,y,z}$ takes into account only dimensions $x$ through $y$, of which exactly $z$ have the value $1$ in the corresponding value in $i$.
Thus $a_{1,K,L}=\sum_{i\in {K\choose L}} f_i(u)$, the original problem we started with.

It's easy to see the following recurrence relation holds:
\begin{eqnarray}
a_{x,y,z}=a_{x,y-1,z} f_0(u_{y}) + a_{x,y-1,z-1} f_1(u_{y}),\\
a_{x,x-1,*}=1
\end{eqnarray}

Using this relation, we can compute $a_{1,K,L}$ in $O(KL)$ time by computing each $a_{1,y,z}$
for $1\leq y\leq K$, $0\leq z\leq L$. In the following subsections we will similarly compute $a_{x,y,z}$ for  $y=K$ and $1\leq x\leq K$ using the following recurrence relation:

\begin{eqnarray}
\label{dp_rr2}
a_{x,y,z}=a_{x+1,y,z} f_0(u_x) + a_{x+1,y,z-1} f_1(u_x),\\
a_{y+1,y,*}=1
\end{eqnarray}

\subsection{Computing the sum in equation \eqref{FWER-const}}
\label{dp_33}
For equation \eqref{FWER-const} we need to compute the following sum:
\begin{equation}
b_k=\sum_{i\in {K\choose L} s.t. i_{min}=k} f_i(u),
\end{equation}
where $i_{min}$ is the minimal index $j$ such that $i_j=0$. We are required to compute $b_k$ for all $1\leq k \leq K$. we will now show that under the i.i.d. assumption they can be computed in $O(KL)$ time.

Note that $i\in{K\choose L} s.t. i_{min}=k$ is a vector such that the first $k-1$ elements are equal to 1, the $k$'th element is zero, and of the remaining $K-k$ elements $L-k+1$ are equal to 1. We will use this observation to rewrite $b_k$:

\begin{equation}
b_k=\left(\prod_{j=1}^{k-1} f_1(u_j)\right) f_0(u_k) a_{k+1,K,L-k+1}
\end{equation}

The prefix  $\prod_{j=1}^{k-1} f_1(u_j)$ can be easily computed for all $1\leq k\leq K$ in $O(K)$ using dynamic programming. The required $a_{k+1,K,L-k+1}$ elements can also be computed using recurrence relation~\ref{dp_rr2} in $O(KL)$ time.

\subsection{Computing the sum in equation \eqref{FDR-const}}

For equation \eqref{FDR-const} we need to compute a different sum of exponential functions:

\begin{equation}
\label{sum_34}
c_k=\sum_{i\in {K\choose L}} f_i(u)r_{ki},
\end{equation}

For computing it, we will use a modified notation for $r_{ki}$, as follows:

\begin{equation*}
r_{x,w,k}=
\left \{
  \begin{array}{ll}
  \frac{k-(x+w)}{k}-\frac{k-1-x}{k-1} & \text{if } k>1,\\
1-w   & \text{if } k=1.
  \end{array}\right.
\end{equation*}

To get the equality $r_{x,w,k}=r_{ki}$ we will set $x=|\{1,\ldots,k-1\}\cap i|$ (with $x=0$ for $k=1$) and $w=i_k$. This new notation is useful since it can be computed without full knowledge of $i$. Only the number of non-zero elements of $i$ in the first $k-1$ dimensions ($x$), and the value of the $k$'th element ($w$) are required.

Now we will define $d_{x,k}$ which is similar to the sum in equation~\ref{sum_34}, but considering only the first $k$ dimensions, and such that $x$ elements of each $i$ are equal to 1:

\begin{equation}
d_{x,k}=\sum_{i\in {k \choose x}} \prod_{j=1}^k f_{i_j}(u_j)r_{ki},
\end{equation}

Note that $r_{ki}$ in the above equation is either $r_{x,0,k}$, if the $x$ non-zero elements of $i$ are within the first $k-1$ dimensions, or $r_{x-1,1,k}$ otherwise. Using $a_{x,y,z}$ from subsection~\ref{dp_basic}, we can rewrite $d_{x,k}$ as follows:
\begin{equation}
d_{x,k}=a_{1,k-1,x} f_0(u_k) r_{x,0,k}+a_{1,k-1,x-1} f_1(u_k) r_{x-1,1,k}
\end{equation}

Thus we can compute $d_{x,k}$ for all $1\leq k\leq K$ and $0\leq x\leq k$ in $O(K^2)$ time.

Finally, the required sum $c_k$ can be rewritten as:
\begin{equation}
c_k=\sum_{x=0}^k d_{x,k} a_{k+1,K,L-x}
\end{equation}

Since we can compute $a_{k+1,K,L-x}$ for all $1\leq k\leq K$, $0\leq x\leq L$ in $O(KL)$ time, we can compute $c_k$ for all $1\leq k\leq K$ in a total of $O(K^2)$ time.

\bibliography{OptimalBib}
\bibliographystyle{apalike}

\end{document}